\documentclass[aps,prl,twocolumn,superscriptaddress,showkeys,floatfix]{revtex4} 
\usepackage{amsmath}
\usepackage{bm}
\usepackage{amssymb}
\usepackage{graphicx}
\usepackage{epstopdf}
\usepackage{bmpsize}
\usepackage{epsfig}
\usepackage{color}
\usepackage{hyperref}
\usepackage{float}
\usepackage{braket}
\usepackage{lipsum}
\usepackage{graphicx}
\usepackage{dcolumn}
\usepackage{bm}
\usepackage{hyperref}
\usepackage[mathlines]{lineno}
\usepackage{amsmath}
\usepackage{amssymb}
\usepackage{graphicx}
\usepackage{epstopdf}
\usepackage{bmpsize}
\usepackage{epsfig}
\usepackage{color}
\usepackage{float}
\usepackage{braket}
\usepackage{lipsum}
\usepackage[toc]{appendix}
\usepackage{amsmath,mathtools,amssymb,tikz}
\usepackage{comment}
\begin{document}
\date{\today}

\preprint{APS/123-QED}


\title{Purcell-enhanced dipolar interactions in nanostructures}

\author{A. Skljarow}
\affiliation{%
 5.Physikalisches Institut and Center for Integrated Quantum Science and Technology, Universität Stuttgart, Pfaffenwaldring 57, 70569 Stuttgart, Germany
}

\author{H. K\"ubler}
\affiliation{%
5.Physikalisches Institut and Center for Integrated Quantum Science and Technology, Universität Stuttgart, Pfaffenwaldring 57, 70569 Stuttgart, Germany
}

\author{C. S. Adams}
\affiliation{%
Joint Quantum Center (JQC) Durham-Newcastle, Department of Physics, Durham University, South Road, Durham, DH3LE, United Kingdom
}

\author{T. Pfau}
\affiliation{%
 5.Physikalisches Institut and Center for Integrated Quantum Science and Technology, Universität Stuttgart, Pfaffenwaldring 57, 70569 Stuttgart, Germany
}

\author{R. L\"ow}
\affiliation{5.Physikalisches Institut and Center for Integrated Quantum Science and Technology, Universität Stuttgart, Pfaffenwaldring 57, 70569 Stuttgart, Germany
}

\author{H. Alaeian}
\affiliation{Elmore Family School of Electrical \& Computer Engineering, Department of Physics \& Astronomy, Purdue Quantum Science \& Engineering Institute, Purdue University, West Lafayette, IN 47907, USA}

\begin{abstract}
Strong light-induced interactions between atoms are known to cause nonlinearities at a few-photon level which are crucial for applications in quantum information processing. Compared to free space, the scattering and the light-induced dipolar interaction of atoms can be enhanced by a dielectric environment. For this \emph{Purcell effect}, either a cavity or a waveguide can be used. Here, we combine the high densities achievable in thermal atomic vapors with an efficient coupling to a slot waveguide. In contrast to free-space interactions, atoms aligned within the slot exhibit repulsive interactions that are further enhanced by a factor of 8 due to the Purcell effect. The corresponding blueshift of the transition frequency of atoms arranged in the essentially one-dimensional geometry vanishes above the saturation, providing a controllable nonlinearity at the few-photon level. The experimental results are in good agreement with Monte-Carlo simulations that include the dielectric environment, dipolar interactions, and motional effects.
The results pave the way towards a robust scalable platform for quantum nonlinear optics and all-optical quantum information processing at room temperature.
\end{abstract}


\maketitle

\begin{figure}[t]
    \centering
    \includegraphics[width = \columnwidth]{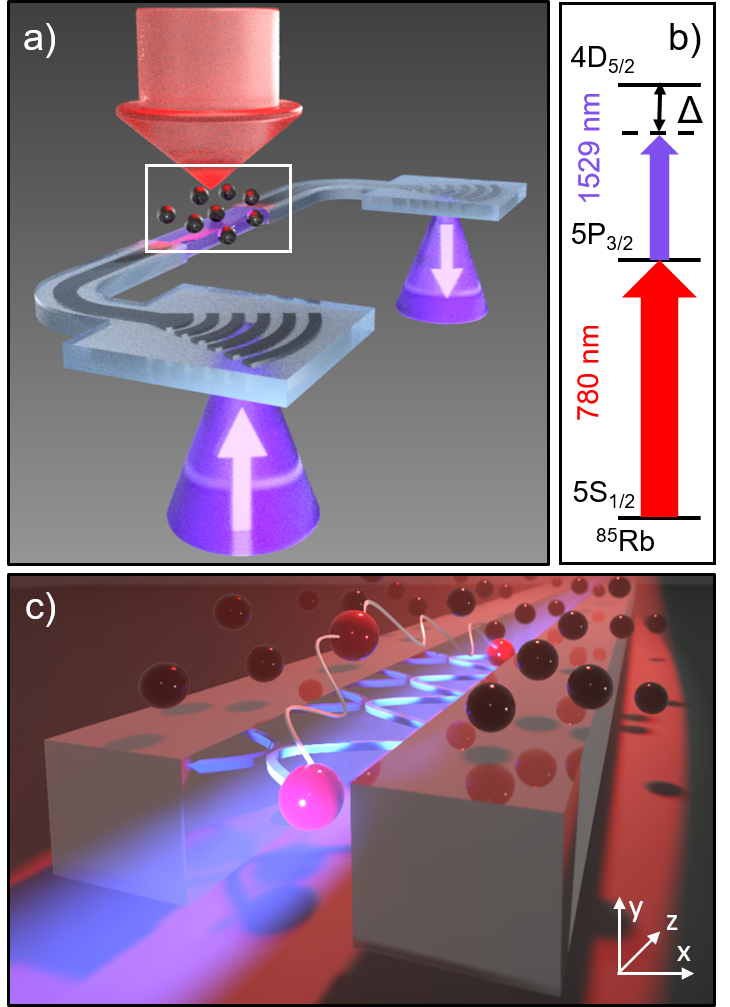}
    \caption{\textbf{Measurement scheme and principles of 1D dipolar interactions}.
    a) Sketch of the atomic cloud interfaced with the nanostructure. The 780~nm pump laser (red) is illuminating the device from the top populating the 5P$_{3/2}$ state. The 1529~nm probe laser is coupled in and out of the slot waveguide via grating couplers. The atom-light interaction takes place in the uncovered slot area in the middle of the device. b) Simplified level scheme of rubidium atoms excited by the pump laser at 780~nm and the probe laser at 1529~nm. c) Magnified image of the interaction region, denoted with the white box in a). The guiding mode (purple) is strongly confined inside the 50~nm slot. Excited atoms (red) can emit a photon either into free space or the waveguide mode. Thicker lines mean stronger interactions mediated via the waveguide mode-coupled photons.}
    \label{fig:fig1}
\end{figure}

\begin{figure*}[t]
   \centering
    \includegraphics[scale = 0.9]{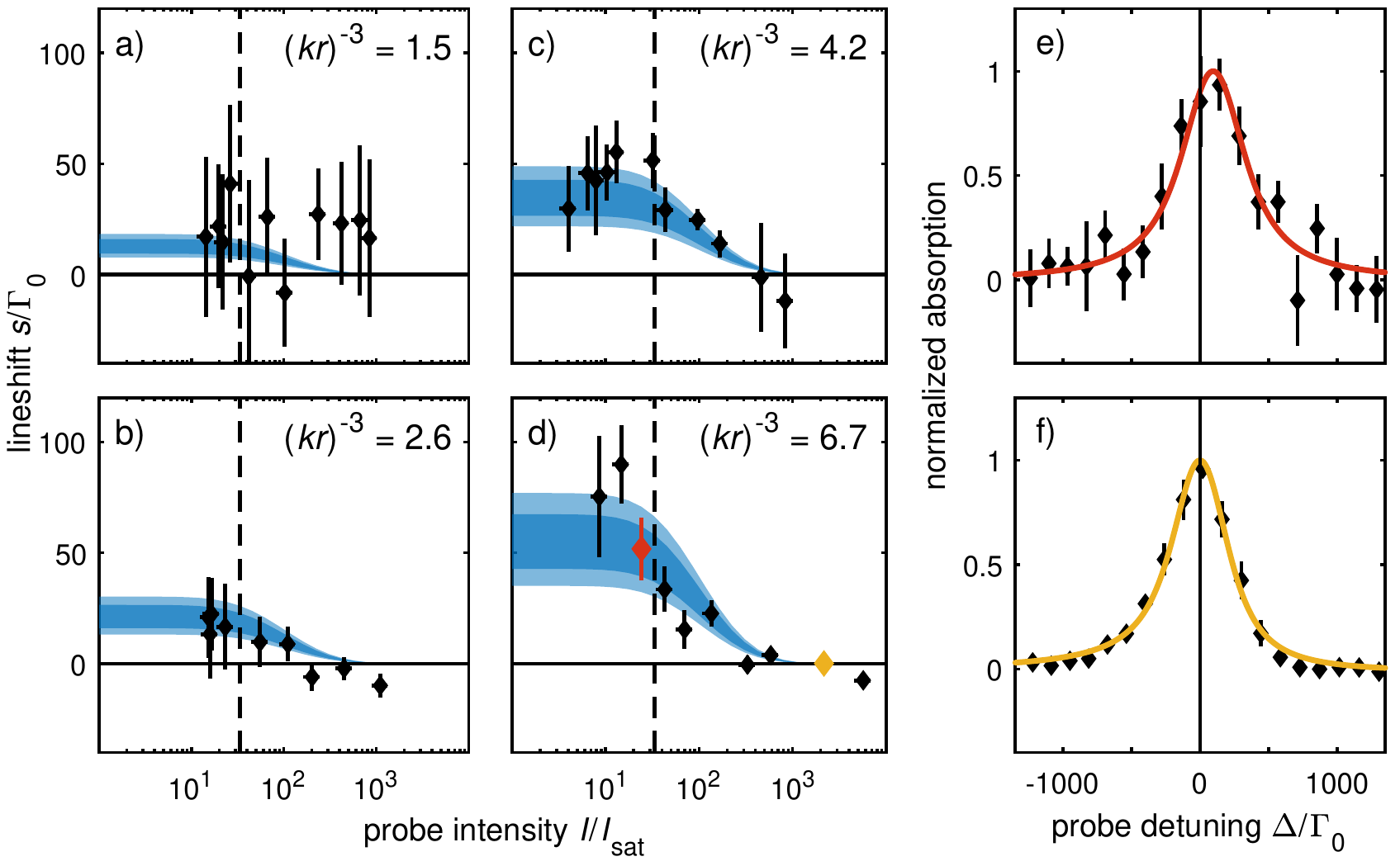}
    \caption{\textbf{Optical nonlinearity and repulsive dipolar interactions as a function of density}.
    a)-d) Transmitted signal shift as a function of probe intensity for different normalized atomic densities $(kr)^{-3}$. We compare the measured lineshifts (black diamonds) with Monte-Carlo simulations (blue). Vertical errorbars are obtained from fit uncertainties of the measured data and horizontal errorbars account for probe power fluctuations during the measurement. For the dark blue shaded area we assume an uncertainty of $\pm$ 5°C in the cell temperature corresponding to a variation of the atomic density. The light blue shaded area stand for a $\pm$ 20$\%$ variation of the Purcell factor as we retrieve this value only from simulations. The horizontal zero line marks the vanishing point of the dipole-dipole shift. The vertical dashed line marks the intensity at which the maximum lineshift is reduced by $1/e^2$ (14$\%$). e)-f) Measured absorption vs probe detuning at $(kr)^{-3} = 6.7$ (black dots) and corresponding fit function. The traces are color-coded with the data points in d) showing the lineshift below (e) and above (f) saturation. Vertical errorbars represent bin standard deviations.}
    \label{fig:fig2}
\end{figure*}

\begin{figure}[t]
    \centering
    \includegraphics[width = 0.8\columnwidth]{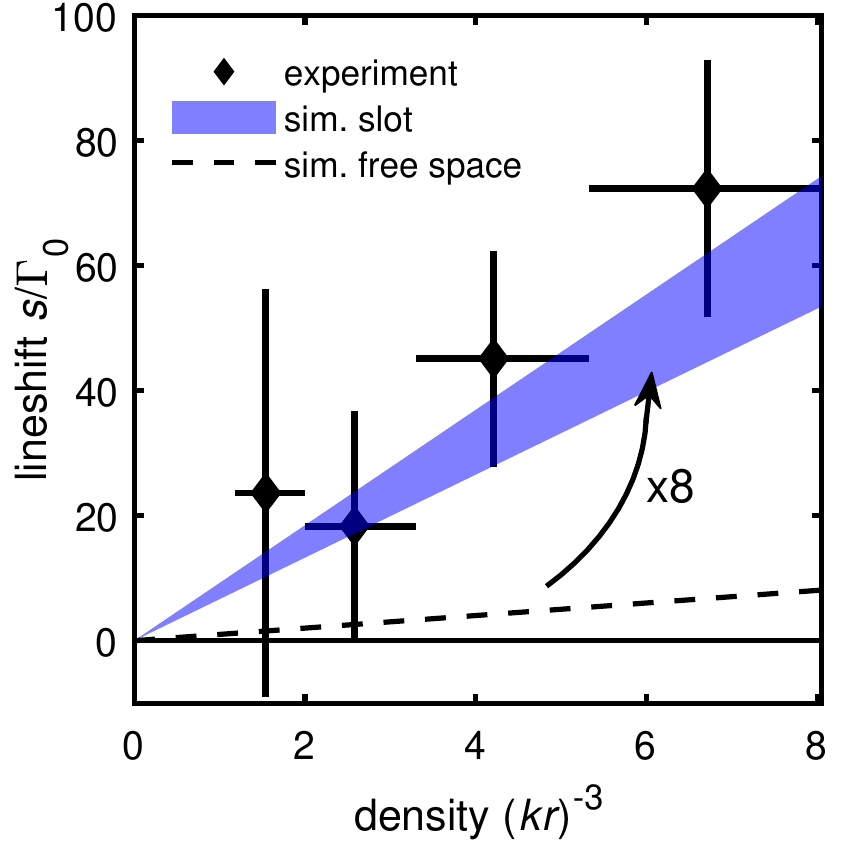}
    \caption{\textbf{Purcell-enhanced interaction}.
    Transmitted signal shift as a function of the normalized density. We plot the averaged lineshift over the data points at weak probe, i.e. lower than the dashed line in Fig.~\ref{fig:fig2} (black diamonds). Horizontal errorbars stem from temperature uncertainties of $\pm$5°C and vertical errorbars are RMS errorbars from the corresponding data points in Fig.~\ref{fig:fig2}. The dashed line depicts the expected free space behaviour while the blue stripe shows the shift obtained from Monte-Carlo simulations including Purcell factor uncertainties (20$\%$) caused by potential fabrication imperfections.}
    \label{fig:fig3}
\end{figure}

\textbf{Introduction -} The theory of atom-light interaction has to be revised as soon as two or more atoms can exchange photons at a non-negligible rate. The corresponding dipole-dipole interaction is relevant in dense media or if a radiation mode is enhanced to couple distant atoms within this mode. The marked differences between such cooperative effects~\cite{coop} and the individual atom were investigated in Dicke's pioneering work, where he introduced the notion of super(sub)-radiant states ~\cite{Dicke1954} leading to extensive studies on ultracold atomic gases in free space ~\cite{coldshift1, coldshift2, coldshift3, Cold_dipole, collective_science, Theory2D}.\newline
Guided modes have the advantage of diffraction-free propagation while providing mode
confinement with an effective area that can be smaller than the free-space diffraction limit. These properties allow the use of waveguides to efficiently interconnect many emitters via a shared mode~\cite{Levy}. The photon exchange between atoms is related to photon exchange between an atom and a cavity or waveguide which are altered by the Purcell effect~\cite{Purcell}. It can modify~\cite{Kimble_Chang} and enhance~\cite{Charles2D} the optical response of a medium. 
Cold atoms trapped and coupled to a nanofiber~\cite{Vetsch2010, NanoFiber} or cavity~\cite{Alligator, Rempe_cavities}, ions inside a cavity~\cite{Walther}, quantum dots coupled to a photonic crystal waveguide~\cite{Arcari2014, Lodahl, Javadi2015}, and organic molecules interfaced with photonic waveguides~\cite{Faez2014,Turschemann2017, Cold_waveguide} are various platforms that benefit from interfacing emitters with nanostructures.
The advantage of thermal atoms over most platforms is that very high densities can be switched on and off on a nanosecond timescale via a process called light induced atomic desporption~\cite{LIAD}. Thermal atoms are  therefore the most flexible nonlinear medium as far as scalability and integrability are concerned. So far, high-density cooperative effects were exclusively studied in thermal 2D systems, e.g., by probing atoms close to the surface via selective reflection spectroscopy~\cite{Ducloy, Gallagher} or in nanometer thin cells~\cite{thincell,thincell1, christaller} where attractive interactions manifest themselves as a redshift and a broadening of the transmission spectrum consistent with the Lorentz-Lorenz scaling~\cite{LL}.\newline
In this letter, we report the observation of Purcell-enhanced cooperative effects in a 1D dense thermal rubidium vapor inside a deep sub-wavelength slot waveguide. We investigate the transition from the many-body regime, where the cooperative effects dominate, to the single-body dynamics by changing the driving field intensity from well below to above saturation. Moreover, we observe the effect of a repulsive all-to-all dipolar interaction mediated via the waveguide mode by measuring the blueshift of the transmission spectrum. We compare our results with a dynamical Monte-Carlo simulation that allows one to include the motion of the atoms, the Casimir-Polder surface effect, the time-dependent interaction among moving atoms via both the vacuum and the waveguide, and the transit broadening due to the inhomogeneity of the waveguide mode hence the driving Rabi frequency.\newline
\textbf{Setup -} 
Our experimental platform consists of thermal Rb vapor immersed in a silicon slot waveguide cell as depicted in Fig.~\ref{fig:fig1}a. The slot waveguide is composed of two ridge waveguides ($w = 300$~nm, $h = 250$~nm, $l = 200~\mu$m) separated by a distance of $g = 50$~nm. This arrangement leads to a strong 1D confinement of the guided mode within the gap 
creating a maximum Purcell factor of $\approx35$ (cf. methods). The whole device is covered by SiO$_2$ except a $L = 200$~$\mu$m-long interaction area. This open region allows for interactions between thermal atoms and the guided mode. In order to excite the atoms, a resonant pump laser at 780~nm, locked to the 5S$_{1/2}$ F = 3 to 5P$_{3/2}$ F' = 4 transition, is focused onto the interaction region from the above. Note that the lifetime in the 5P$_{3/2}$ state (27 ns ~\cite{Steck}) is much longer than the transient time set by the waveguide geometry. Therefore, the 780 nm pump laser can be used to vary the density of atoms excited to the intermediate state. Since the probing of atoms only happens in about 1~ns in the vicinity of the slot waveguide, we do not observe noticeable optical pumping. A probe laser at $\lambda = 1529$~nm excites the waveguide mode via a grating coupler. Atoms are further excited to the 4D$_{5/2}$ state by the guided probe laser that is scanned over all 4D hyperfine states with a detuning $\Delta$ from resonance (see Fig.~\ref{fig:fig1}b).
Each atom can emit a photon either into free space (thin wavy line) or into the slot mode (thick wavy line), with probabilities depending on the position of the atoms, as depicted with wavy lines for one particular atom (red sphere) at the front of the waveguide in Fig.~\ref{fig:fig1}c. Hence, the transmitted probe contains information about the dipole-dipole interaction inside the slot (cf. methods).\newline
\textbf{Results and discussion -}
In Fig.~\ref{fig:fig2}a-d) we show the lineshift of the slot transmission signal for a broad range of probe intensities below and above saturation, and at 4 different atomic densities. The black data points represent the measured shift while the blue stripes represent the theoretical predictions. These shifts are obtained from the transmission spectrum by fitting a Fano lineshape convoluted with a Voigt profile~\cite{FanoVoigt} (cf. methods). The additional Fano profile helps to cope with interferences occurring at high intensities. We use this fit function to capture the lineshape consistently in the weak probe regime (Fig.~\ref{fig:fig2}e) and in saturation (Fig.~\ref{fig:fig2}f). From the density $n$ we determine the mean inter-atomic distance $r = (4\pi/3\times n)^{-1/3}$ and define a normalized density $(kr)^{-3}$ with the free-space wavevector $k=2\pi/\lambda$ of the probe laser. The effect of interactions is negligible at low densities (Fig.~\ref{fig:fig2}a) and becomes more prominent as the density increases (Fig.~\ref{fig:fig2}b-d). We observe a blueshift of the transmission spectra in the weak-probe limit that is continuously vanishing with increasing probe intensity as expected. Due to the decoherence introduced by the short transient time and Doppler broadening on the order of 1~GHz, the probe intensity must be roughly 500 times larger than the natural saturation intensity of $I_\text{sat} \approx 0.14$~mW/cm$^2$ in order to saturate the transition. The blueshift can be explained by considering two interacting dipoles which experience an attractive interaction (redshift) in a head-to-tail arrangement and a repulsive interaction (blueshift) for a side-by-side configuration. Due to the strong horizontal ($g<\lambda/30$) and vertical ($h<\lambda/6$) confinement of the slot, excited atoms are polarized perpendicular to the waveguide axis and render a pseudo 1D chain along the waveguide. Thus, the blueshift arises from a dominantly side-by-side alignment of the dipoles within the slot.
We can gradually tune the transition from this interacting many-body regime to a non-interacting one by increasing the probe intensity~\cite{coldshift2}. As the intensity increases towards saturation the photon exchange between excited atoms vanishes leading to the suppression of their interaction.
Nonlinearites on a single-photon level, meaning a $\pi$-phase shift induced by one photon, are a sought-after goal in quantum optics~\cite{pi_phase}.
In order to describe the nonlinearity we estimate the length-independent single-photon phase shift $\phi/´(k_\text{eff}L) = n_2I_\text{photon}$, with the effective wavevector $k_\text{eff} = n_\text{eff}k$ obtained from the effective mode index $n_\text{eff} = 2.53$ at 1529~nm, the Kerr coefficient $n_2$ and the single-photon intensity $I_\text{photon}$ (cf. methods). For our largest measured nonlinearity at a probe intensity of $I=22I_\text{sat}$ and at a density of $(kr)^{-3} = 6.7$ (see Fig.~\ref{fig:fig2}d) we obtain $n_2I_\text{photon} = 3.30\times10^{-5}$ ($\phi = 0.07$~rad) with a Kerr coefficient of $n_2 = 1.98\times10^{-7}$~m$^2$/W. Our rough estimate of the phase shift per photon is about 3 orders of magnitude larger than the reported values for an ultracold sodium BEC $n_2I_\text{photon}(\text{Na}) = 4.47\times10^{-8}$ with a Kerr coefficient of $n_2 = 1.80\times 10^{-5}$~m$^2$/W~\cite{c17ms} and ultracold rubidium $n_2I_\text{photon}(\text{Rb}) = 1.77\times10^{-9}$~\cite{EIT_RbXPS}. Although, our attainable Kerr coefficient is rather small compared to ultracold systems, we benefit from a strong confinement of the guided mode leading to a large single-photon intensity.\\
In Fig.~\ref{fig:fig2}e-f we show the normalized probe absorption as a function of detuning in units of the natural decay rate $\Gamma_0/(2\pi) = 1.89$~MHz, for the weak and strong probe case, respectively. The black dots show the measured spectrum and solid lines correspond to the fits. The black vertical line denotes the position of the cyclic transition (5S$_{1/2}$ F = 3$\rightarrow$ 5P$_{3/2}$ F = 4 $\rightarrow$ 4D$_{5/2}$ F = 5) of the hyperfine reference spectrum (cf. methods).\newline
Within the investigated density regime $1<(kr)^{-3}<7$ we expect a linear dependency of the lineshift with density as summarized in Fig.~\ref{fig:fig3}. There, the measured data (black dots) is obtained by averaging over the data points in the weak probe regime in Fig.~\ref{fig:fig2}a-d. We choose to cut off the weak probe region at an intensity at which the dipolar lineshift is reduced by $1/e^2$ (dashed line in Fig.~\ref{fig:fig2}). Overall the Purcell effect leads to an average factor of roughly 8 enhancement of the dipole-dipole interactions (black arrow in Fig.~\ref{fig:fig3}). Note that while the local Purcell factor can be up to 35 right at the center of the gap (cf. methods), the flying atoms interact with the full mode which leads to an averaging of this effect due to their stochastic motions and positions. The dashed line shows the expected lineshift without the Purcell enhancement for comparison.
\newline
\noindent
\textbf{Theory -} 
We investigate the system with a dynamical Monte-Carlo method, that has been developed in our group previously \cite{Ralf_slot}, to explore the atom-light interaction. In our recent work~\cite{Artur} we studied all relevant single-body effects theoretically and experimentally for a ridge waveguide in the dilute regime and found good agreement between the measurement and simulation. To describe the atom-atom interactions in this work we simplify the atoms to two levels (5P$\rightarrow$4D).
We perform time-resolved simulations of an atomic ensemble each with random velocity $v_\text{m}$ according to the Maxwell-Boltzmann distribution at location $\textbf{a}_m(t)$ starting from the ground state at $t=0$, i.e. $\braket{\hat{\sigma}_\text{gg}^{(m)}(0)}= 1$. Atoms obtain a Doppler shift $\Delta_\text{D} = n_\text{eff}kv$ along the waveguide. At every time step, the m$^{th}$ atom experiences an effective Rabi frequency $\Omega_\text{eff}^{m}$ as 
\begin{equation}
    \Omega_\text{eff}^m = \Omega_0^m+3\pi\sum_{m\neq m'}\left(G^{m,m'}_\text{FS}+G^{m,m'}_\text{WG}\right)\braket{\hat{\sigma}^{(m')}_\text{ge}},
    \label{Oeff}
\end{equation}
where in the above equation, $\Omega_0^m$ is the Rabi frequency due to the waveguide mode and the second term is the collective mean-field of all other atoms $m'$ modifying the m$^{th}$ atom local field~\cite{Chang_manybody}. Furthermore, $\braket{\hat{\sigma}^{(m')}_\text{ge}}$ is the m$'^{th}$ atom coherence and $G_\text{FS}^{m,m'}$ and $G_\text{WG}^{m,m'}$ are the free-space and waveguide Green's functions, respectively, reading as
\begin{equation}
    G_\text{FS}^{m,m'}= \left(\bar{\bar{I}}+\frac{1}{k^2}\nabla\nabla\right)\frac{e^{ik|\textbf{a}_m-\textbf{a}_{m'}|}}{4\pi k|\textbf{a}_m-\textbf{a}_{m'}|},
    \label{GFS}
\end{equation}
and
\begin{equation}
    G_{\text{WG}}^{{m,m'}}= i\frac{\Gamma_\text{WG}^{m,m'}}{\Gamma_0}e^{in_\text{eff}k|z_m-z_{m'}|},
    \label{GWG}
\end{equation}
where $\Gamma_\text{WG}$ stands for the position dependent coupling rate to the waveguide and $z_m$ for the location of the atoms in the direction of the probe propagation (see Fig.~\ref{fig:fig1}c). It can be shown that the Purcell factor at each point is related to the waveguide Green's function (cf. methods). This approach ignores all correlations which is a valid assumption since they are quickly lost due to transient interactions on a timescale of $\approx$1~ns~\cite{JavanienLLBec}. Within the validity range of the Born approximation it is reasonable to add the free-space and waveguide Green's functions to find the effective total Green's function.
A detailed description of this model can be found in the methods section.
The simulated densities of the N-atom ensemble is calculated via $n=N/V$ where the probe volume $V$ is chosen such that the guiding mode vanishes at the edges. At any collision with the slot waveguide or the boundaries of the probe volume atoms are de-excited to the ground state and the corresponding coherence is lost. In order to obtain a spectrum we average the excited state population $\braket{\hat{\sigma}^{(m)}_\text{ee}(t)}$, obtained from solving the generalized optical Bloch equations at each time step, over all atoms and all time steps at different detunings $\Delta$. The traces in Fig.~\ref{fig:fig2} are obtained by calculating a spectrum for different driving field strengths $\Omega_0$. The blue stripe in Fig.~\ref{fig:fig3} is obtained from a simulation in the weak-probe regime at different densities.
For comparison, the shift resulting from the free-space interaction, i.e. ignoring the Purcell enhancement ($\Gamma_\textrm{WG} = 0$), is plotted in Fig.~\ref{fig:fig3} in dashed lines.
We find good agreement between the theory and measurement without any free parameters.\newline
\textbf{Conclusion - } 
This work reports the observation of a Purcell-enhanced dipole-dipole interaction in a dense thermal rubidium vapor and a 1D arrangement set by a sub-wavelength slot waveguide. By preparing a one dimensional chain of side-by-side atoms we mostly harvest the repulsive dipolar interaction which leads to a blueshift of the transmission spectrum. 
Due to an all-to-all interaction between the atoms inside the slot it is possible to surpass the $1/r^3$ length scale of the free-space interaction which results in a strong enhancement of the repulsion.
Benefitting from the miniaturized system, we have been able to achieve highly nonlinear behaviour with an estimated number of 45 photons required for a phase shift of $\pi$ out of which $\approx8\%$ are absorbed on resonance.
Exclusive to our scalable and integrated platform we can use in future the effect of light induced atomic desorption in order to switch the density by 2 orders of magnitude within a nanosecond timescale ~\cite{christaller}. 
Via coupling to the guided mode it is also possible to increase the effective optical depth without changing the actual density of emitters~\cite{Chang_OD} useful for memory applications~\cite{Antoine_memory}. The ability to enhance or suppress the emission into selected modes is highly interesting in topological quantum systems. It gives one the opportunity to unite chiral~\cite{Buchler_Topo} and topological~\cite{Topo_Lukin} structures with the inherent strong nonlinearity of atoms.
Equally, methods that increase the Purcell factor to a regime where the shift overcomes the broadening of the system could lead to a novel type of nonlinearity at a single-photon level~\cite{WG_Blockade}.
This enhancement is not only limited to atomic systems and can be further extended to photon exchange between any quantum emitter as long as all contributing particles interact with each other resonantly~\cite{Foester}.
\newline
\textbf{Acknowledgements - } H.A. would like to thank D. E. Chang for the insightful discussions at the early stages of this project. We gratefully acknowledge the financial support by Baden-W\"urttemberg Stiftung via Grant BWST$\_$ISF2019-017 under the program Internationale Spitzenforschung. H.A. acknowledges the financial support from Integrated Quantum Science and Technology (IQST) center, Baden-W\"urttemberg Stiftung, and Purdue University Startup grant. C.S.A. is supported by UK EPSRC Grant Ref. No.  EP/V030280/1.\\

\bibliography{Ref}

\begin{thebibliography}{48}
\expandafter\ifx\csname natexlab\endcsname\relax\def\natexlab#1{#1}\fi
\expandafter\ifx\csname bibnamefont\endcsname\relax
  \def\bibnamefont#1{#1}\fi
\expandafter\ifx\csname bibfnamefont\endcsname\relax
  \def\bibfnamefont#1{#1}\fi
\expandafter\ifx\csname citenamefont\endcsname\relax
  \def\citenamefont#1{#1}\fi
\expandafter\ifx\csname url\endcsname\relax
  \def\url#1{\texttt{#1}}\fi
\expandafter\ifx\csname urlprefix\endcsname\relax\def\urlprefix{URL }\fi
\providecommand{\bibinfo}[2]{#2}
\providecommand{\eprint}[2][]{\url{#2}}

\bibitem[{\citenamefont{Reitz et~al.}(2021)\citenamefont{Reitz, Sommer, and
  Genes}}]{coop}
\bibinfo{author}{\bibfnamefont{M.}~\bibnamefont{Reitz}},
  \bibinfo{author}{\bibfnamefont{C.}~\bibnamefont{Sommer}}, \bibnamefont{and}
  \bibinfo{author}{\bibfnamefont{C.}~\bibnamefont{Genes}},
  \bibinfo{journal}{arXiv:2107.02674}  (\bibinfo{year}{2021}).

\bibitem[{\citenamefont{Dicke}(1954)}]{Dicke1954}
\bibinfo{author}{\bibfnamefont{R.~H.} \bibnamefont{Dicke}},
  \bibinfo{journal}{Phys. Rev.} \textbf{\bibinfo{volume}{93}},
  \bibinfo{pages}{99} (\bibinfo{year}{1954}).

\bibitem[{\citenamefont{Roof et~al.}(2016)\citenamefont{Roof, Kemp, Havey, and
  Sokolov}}]{coldshift1}
\bibinfo{author}{\bibfnamefont{S.~J.} \bibnamefont{Roof}},
  \bibinfo{author}{\bibfnamefont{K.~J.} \bibnamefont{Kemp}},
  \bibinfo{author}{\bibfnamefont{M.~D.} \bibnamefont{Havey}}, \bibnamefont{and}
  \bibinfo{author}{\bibfnamefont{I.~M.} \bibnamefont{Sokolov}},
  \bibinfo{journal}{Phys. Rev. Lett.} \textbf{\bibinfo{volume}{117}},
  \bibinfo{pages}{073003} (\bibinfo{year}{2016}).

\bibitem[{\citenamefont{Glicenstein et~al.}(2020)}]{coldshift2}
\bibinfo{author}{\bibfnamefont{A.}~\bibnamefont{Glicenstein}}
  \bibnamefont{et~al.}, \bibinfo{journal}{Phys. Rev. Lett.}
  \textbf{\bibinfo{volume}{124}}, \bibinfo{pages}{253602}
  (\bibinfo{year}{2020}).

\bibitem[{\citenamefont{Corman et~al.}(2017)}]{coldshift3}
\bibinfo{author}{\bibfnamefont{L.}~\bibnamefont{Corman}} \bibnamefont{et~al.},
  \bibinfo{journal}{Phys. Rev. A} \textbf{\bibinfo{volume}{96}},
  \bibinfo{pages}{053629} (\bibinfo{year}{2017}).

\bibitem[{\citenamefont{Guo et~al.}(2019)}]{Cold_dipole}
\bibinfo{author}{\bibfnamefont{M.}~\bibnamefont{Guo}} \bibnamefont{et~al.},
  \bibinfo{journal}{Nature} \textbf{\bibinfo{volume}{574}},
  \bibinfo{pages}{386} (\bibinfo{year}{2019}).

\bibitem[{\citenamefont{Röhlsberger et~al.}(2010)\citenamefont{Röhlsberger,
  Schlage, Sahoo, Couet, and Rüffer}}]{collective_science}
\bibinfo{author}{\bibfnamefont{R.}~\bibnamefont{Röhlsberger}},
  \bibinfo{author}{\bibfnamefont{K.}~\bibnamefont{Schlage}},
  \bibinfo{author}{\bibfnamefont{B.}~\bibnamefont{Sahoo}},
  \bibinfo{author}{\bibfnamefont{S.}~\bibnamefont{Couet}}, \bibnamefont{and}
  \bibinfo{author}{\bibfnamefont{R.}~\bibnamefont{Rüffer}},
  \bibinfo{journal}{Science} \textbf{\bibinfo{volume}{328}},
  \bibinfo{pages}{1248} (\bibinfo{year}{2010}).

\bibitem[{\citenamefont{Bettles et~al.}(2020)\citenamefont{Bettles, Lee,
  Gardiner, and Ruostekoski}}]{Theory2D}
\bibinfo{author}{\bibfnamefont{R.~J.} \bibnamefont{Bettles}},
  \bibinfo{author}{\bibfnamefont{M.~D.} \bibnamefont{Lee}},
  \bibinfo{author}{\bibfnamefont{S.~A.} \bibnamefont{Gardiner}},
  \bibnamefont{and}
  \bibinfo{author}{\bibfnamefont{J.}~\bibnamefont{Ruostekoski}},
  \bibinfo{journal}{Communications Physics} \textbf{\bibinfo{volume}{3}},
  \bibinfo{pages}{141} (\bibinfo{year}{2020}).

\bibitem[{\citenamefont{Stern et~al.}(2013)\citenamefont{Stern, Desiatov,
  Goykhman, and Levy}}]{Levy}
\bibinfo{author}{\bibfnamefont{L.}~\bibnamefont{Stern}},
  \bibinfo{author}{\bibfnamefont{B.}~\bibnamefont{Desiatov}},
  \bibinfo{author}{\bibfnamefont{I.}~\bibnamefont{Goykhman}}, \bibnamefont{and}
  \bibinfo{author}{\bibfnamefont{U.}~\bibnamefont{Levy}},
  \bibinfo{journal}{Nature Communications} \textbf{\bibinfo{volume}{4}},
  \bibinfo{pages}{1548} (\bibinfo{year}{2013}).

\bibitem[{\citenamefont{Purcell et~al.}(1946)\citenamefont{Purcell, Torrey, and
  Pound}}]{Purcell}
\bibinfo{author}{\bibfnamefont{E.~M.} \bibnamefont{Purcell}},
  \bibinfo{author}{\bibfnamefont{H.~C.} \bibnamefont{Torrey}},
  \bibnamefont{and} \bibinfo{author}{\bibfnamefont{R.~V.} \bibnamefont{Pound}},
  \bibinfo{journal}{Phys. Rev.} \textbf{\bibinfo{volume}{69}},
  \bibinfo{pages}{37} (\bibinfo{year}{1946}).

\bibitem[{\citenamefont{Asenjo-Garcia
  et~al.}(2019{\natexlab{a}})\citenamefont{Asenjo-Garcia, Kimble, and
  Chang}}]{Kimble_Chang}
\bibinfo{author}{\bibfnamefont{A.}~\bibnamefont{Asenjo-Garcia}},
  \bibinfo{author}{\bibfnamefont{H.~J.} \bibnamefont{Kimble}},
  \bibnamefont{and} \bibinfo{author}{\bibfnamefont{D.~E.} \bibnamefont{Chang}},
  \bibinfo{journal}{Proceedings of the National Academy of Sciences}
  \textbf{\bibinfo{volume}{116}}, \bibinfo{pages}{25503}
  (\bibinfo{year}{2019}{\natexlab{a}}).

\bibitem[{\citenamefont{Bettles et~al.}(2016)\citenamefont{Bettles, Gardiner,
  and Adams}}]{Charles2D}
\bibinfo{author}{\bibfnamefont{R.~J.} \bibnamefont{Bettles}},
  \bibinfo{author}{\bibfnamefont{S.~A.} \bibnamefont{Gardiner}},
  \bibnamefont{and} \bibinfo{author}{\bibfnamefont{C.~S.} \bibnamefont{Adams}},
  \bibinfo{journal}{Phys. Rev. Lett.} \textbf{\bibinfo{volume}{116}},
  \bibinfo{pages}{103602} (\bibinfo{year}{2016}).

\bibitem[{\citenamefont{Vetsch et~al.}(2010)}]{Vetsch2010}
\bibinfo{author}{\bibfnamefont{E.}~\bibnamefont{Vetsch}} \bibnamefont{et~al.},
  \bibinfo{journal}{Phys. Rev. Lett.} \textbf{\bibinfo{volume}{104}},
  \bibinfo{pages}{203603} (\bibinfo{year}{2010}).

\bibitem[{\citenamefont{Prasad et~al.}(2020)}]{NanoFiber}
\bibinfo{author}{\bibfnamefont{A.~S.} \bibnamefont{Prasad}}
  \bibnamefont{et~al.}, \bibinfo{journal}{Nature Photonics}
  \textbf{\bibinfo{volume}{14}}, \bibinfo{pages}{719} (\bibinfo{year}{2020}).

\bibitem[{\citenamefont{Hood et~al.}(2016)}]{Alligator}
\bibinfo{author}{\bibfnamefont{J.~D.} \bibnamefont{Hood}} \bibnamefont{et~al.},
  \bibinfo{journal}{Proceedings of the National Academy of Sciences}
  \textbf{\bibinfo{volume}{113}}, \bibinfo{pages}{10507}
  (\bibinfo{year}{2016}).

\bibitem[{\citenamefont{Ritter et~al.}(2012)}]{Rempe_cavities}
\bibinfo{author}{\bibfnamefont{S.}~\bibnamefont{Ritter}} \bibnamefont{et~al.},
  \bibinfo{journal}{Nature} \textbf{\bibinfo{volume}{484}},
  \bibinfo{pages}{195} (\bibinfo{year}{2012}).

\bibitem[{\citenamefont{Keller et~al.}(2004)\citenamefont{Keller, Lange,
  Hayasaka, Lange, and Walther}}]{Walther}
\bibinfo{author}{\bibfnamefont{M.}~\bibnamefont{Keller}},
  \bibinfo{author}{\bibfnamefont{B.}~\bibnamefont{Lange}},
  \bibinfo{author}{\bibfnamefont{K.}~\bibnamefont{Hayasaka}},
  \bibinfo{author}{\bibfnamefont{W.}~\bibnamefont{Lange}}, \bibnamefont{and}
  \bibinfo{author}{\bibfnamefont{H.}~\bibnamefont{Walther}},
  \bibinfo{journal}{Nature} \textbf{\bibinfo{volume}{431}},
  \bibinfo{pages}{1075} (\bibinfo{year}{2004}).

\bibitem[{\citenamefont{Arcari et~al.}(2014)}]{Arcari2014}
\bibinfo{author}{\bibfnamefont{M.}~\bibnamefont{Arcari}} \bibnamefont{et~al.},
  \bibinfo{journal}{Phys. Rev. Lett.} \textbf{\bibinfo{volume}{113}},
  \bibinfo{pages}{093603} (\bibinfo{year}{2014}).

\bibitem[{\citenamefont{Appel et~al.}(2021)}]{Lodahl}
\bibinfo{author}{\bibfnamefont{M.}~\bibnamefont{Appel}} \bibnamefont{et~al.},
  \bibinfo{journal}{Phys. Rev. Lett.} \textbf{\bibinfo{volume}{126}},
  \bibinfo{pages}{013602} (\bibinfo{year}{2021}).

\bibitem[{\citenamefont{Javadi et~al.}(2015)}]{Javadi2015}
\bibinfo{author}{\bibfnamefont{A.}~\bibnamefont{Javadi}} \bibnamefont{et~al.},
  \bibinfo{journal}{Nature Communications} \textbf{\bibinfo{volume}{6}},
  \bibinfo{pages}{8655} (\bibinfo{year}{2015}).

\bibitem[{\citenamefont{Faez et~al.}(2014)}]{Faez2014}
\bibinfo{author}{\bibfnamefont{S.}~\bibnamefont{Faez}} \bibnamefont{et~al.},
  \bibinfo{journal}{Phys. Rev. Lett.} \textbf{\bibinfo{volume}{113}},
  \bibinfo{pages}{213601} (\bibinfo{year}{2014}).

\bibitem[{\citenamefont{T\"urschmann et~al.}(2017)}]{Turschemann2017}
\bibinfo{author}{\bibfnamefont{P.}~\bibnamefont{T\"urschmann}}
  \bibnamefont{et~al.}, \bibinfo{journal}{Nano Letters}
  \textbf{\bibinfo{volume}{17}}, \bibinfo{pages}{4941} (\bibinfo{year}{2017}).

\bibitem[{\citenamefont{Corzo et~al.}(2019)}]{Cold_waveguide}
\bibinfo{author}{\bibfnamefont{N.~V.} \bibnamefont{Corzo}}
  \bibnamefont{et~al.}, \bibinfo{journal}{Nature}
  \textbf{\bibinfo{volume}{566}}, \bibinfo{pages}{359} (\bibinfo{year}{2019}).

\bibitem[{\citenamefont{Meucci et~al.}(1994)\citenamefont{Meucci, Mariotti,
  Bicchi, Marinelli, and Moi}}]{LIAD}
\bibinfo{author}{\bibfnamefont{M.}~\bibnamefont{Meucci}},
  \bibinfo{author}{\bibfnamefont{E.}~\bibnamefont{Mariotti}},
  \bibinfo{author}{\bibfnamefont{P.}~\bibnamefont{Bicchi}},
  \bibinfo{author}{\bibfnamefont{C.}~\bibnamefont{Marinelli}},
  \bibnamefont{and} \bibinfo{author}{\bibfnamefont{L.}~\bibnamefont{Moi}},
  \bibinfo{journal}{EPL} \textbf{\bibinfo{volume}{25}}, \bibinfo{pages}{639}
  (\bibinfo{year}{1994}).

\bibitem[{\citenamefont{Nienhuis et~al.}(1988)\citenamefont{Nienhuis, Schuller,
  and Ducloy}}]{Ducloy}
\bibinfo{author}{\bibfnamefont{G.}~\bibnamefont{Nienhuis}},
  \bibinfo{author}{\bibfnamefont{F.}~\bibnamefont{Schuller}}, \bibnamefont{and}
  \bibinfo{author}{\bibfnamefont{M.}~\bibnamefont{Ducloy}},
  \bibinfo{journal}{Phys. Rev. A} \textbf{\bibinfo{volume}{38}},
  \bibinfo{pages}{5197} (\bibinfo{year}{1988}).

\bibitem[{\citenamefont{Guo et~al.}(1996)\citenamefont{Guo, Cooper, and
  Gallagher}}]{Gallagher}
\bibinfo{author}{\bibfnamefont{J.}~\bibnamefont{Guo}},
  \bibinfo{author}{\bibfnamefont{J.}~\bibnamefont{Cooper}}, \bibnamefont{and}
  \bibinfo{author}{\bibfnamefont{A.}~\bibnamefont{Gallagher}},
  \bibinfo{journal}{Phys. Rev. A} \textbf{\bibinfo{volume}{53}},
  \bibinfo{pages}{1130} (\bibinfo{year}{1996}).

\bibitem[{\citenamefont{Peyrot et~al.}(2018)}]{thincell}
\bibinfo{author}{\bibfnamefont{T.}~\bibnamefont{Peyrot}} \bibnamefont{et~al.},
  \bibinfo{journal}{Phys. Rev. Lett.} \textbf{\bibinfo{volume}{120}},
  \bibinfo{pages}{243401} (\bibinfo{year}{2018}).

\bibitem[{\citenamefont{Whittaker et~al.}(2014)}]{thincell1}
\bibinfo{author}{\bibfnamefont{K.~A.} \bibnamefont{Whittaker}}
  \bibnamefont{et~al.}, \bibinfo{journal}{Phys. Rev. Lett.}
  \textbf{\bibinfo{volume}{112}}, \bibinfo{pages}{253201}
  (\bibinfo{year}{2014}).

\bibitem[{\citenamefont{Christaller et~al.}(2021)}]{christaller}
\bibinfo{author}{\bibfnamefont{F.}~\bibnamefont{Christaller}}
  \bibnamefont{et~al.}, \bibinfo{journal}{arXiv:2110.00437}
  (\bibinfo{year}{2021}).

\bibitem[{\citenamefont{Javanainen et~al.}(2014)\citenamefont{Javanainen,
  Ruostekoski, Li, and Yoo}}]{LL}
\bibinfo{author}{\bibfnamefont{J.}~\bibnamefont{Javanainen}},
  \bibinfo{author}{\bibfnamefont{J.}~\bibnamefont{Ruostekoski}},
  \bibinfo{author}{\bibfnamefont{Y.}~\bibnamefont{Li}}, \bibnamefont{and}
  \bibinfo{author}{\bibfnamefont{S.-M.} \bibnamefont{Yoo}},
  \bibinfo{journal}{Phys. Rev. Lett.} \textbf{\bibinfo{volume}{112}},
  \bibinfo{pages}{113603} (\bibinfo{year}{2014}).

\bibitem[{\citenamefont{Steck}(2019)}]{Steck}
\bibinfo{author}{\bibfnamefont{D.~A.} \bibnamefont{Steck}},
  \emph{\bibinfo{title}{Rubidium 85 {D} line data}},
  \bibinfo{howpublished}{available online at http://steck.us/alkalidata}
  (\bibinfo{year}{2019}).

\bibitem[{\citenamefont{Schippers}(2018)}]{FanoVoigt}
\bibinfo{author}{\bibfnamefont{S.}~\bibnamefont{Schippers}},
  \bibinfo{journal}{Journal of Quantitative Spectroscopy and Radiative
  Transfer} \textbf{\bibinfo{volume}{219}}, \bibinfo{pages}{33}
  (\bibinfo{year}{2018}).

\bibitem[{\citenamefont{Vaneecloo et~al.}(2021)\citenamefont{Vaneecloo, Garcia,
  and Ourjoumtsev}}]{pi_phase}
\bibinfo{author}{\bibfnamefont{J.}~\bibnamefont{Vaneecloo}},
  \bibinfo{author}{\bibfnamefont{S.}~\bibnamefont{Garcia}}, \bibnamefont{and}
  \bibinfo{author}{\bibfnamefont{A.}~\bibnamefont{Ourjoumtsev}},
  \emph{\bibinfo{title}{An intracavity rydberg superatom for optical quantum
  engineering: Coherent control, single-shot detection and optical $\pi$ phase
  shift}} (\bibinfo{year}{2021}).

\bibitem[{\citenamefont{Hau et~al.}(1999)\citenamefont{Hau, Harris, Dutton, and
  Behroozi}}]{c17ms}
\bibinfo{author}{\bibfnamefont{L.~V.} \bibnamefont{Hau}},
  \bibinfo{author}{\bibfnamefont{S.~E.} \bibnamefont{Harris}},
  \bibinfo{author}{\bibfnamefont{Z.}~\bibnamefont{Dutton}}, \bibnamefont{and}
  \bibinfo{author}{\bibfnamefont{C.~H.} \bibnamefont{Behroozi}},
  \bibinfo{journal}{Nature} \textbf{\bibinfo{volume}{397}},
  \bibinfo{pages}{594} (\bibinfo{year}{1999}).

\bibitem[{\citenamefont{Feizpour et~al.}(2015)\citenamefont{Feizpour, Hallaji,
  Dmochowski, and Steinberg}}]{EIT_RbXPS}
\bibinfo{author}{\bibfnamefont{A.}~\bibnamefont{Feizpour}},
  \bibinfo{author}{\bibfnamefont{M.}~\bibnamefont{Hallaji}},
  \bibinfo{author}{\bibfnamefont{G.}~\bibnamefont{Dmochowski}},
  \bibnamefont{and} \bibinfo{author}{\bibfnamefont{A.~M.}
  \bibnamefont{Steinberg}}, \bibinfo{journal}{Nature Physics}
  \textbf{\bibinfo{volume}{11}}, \bibinfo{pages}{905} (\bibinfo{year}{2015}).

\bibitem[{\citenamefont{Ritter et~al.}(2018)}]{Ralf_slot}
\bibinfo{author}{\bibfnamefont{R.}~\bibnamefont{Ritter}} \bibnamefont{et~al.},
  \bibinfo{journal}{Phys. Rev. X} \textbf{\bibinfo{volume}{8}},
  \bibinfo{pages}{021032} (\bibinfo{year}{2018}).

\bibitem[{\citenamefont{Skljarow et~al.}(2020)}]{Artur}
\bibinfo{author}{\bibfnamefont{A.}~\bibnamefont{Skljarow}}
  \bibnamefont{et~al.}, \bibinfo{journal}{Opt. Express}
  \textbf{\bibinfo{volume}{28}}, \bibinfo{pages}{19593} (\bibinfo{year}{2020}).

\bibitem[{\citenamefont{Fayard et~al.}(2021)\citenamefont{Fayard, Henriet,
  Asenjo-Garcia, and Chang}}]{Chang_manybody}
\bibinfo{author}{\bibfnamefont{N.}~\bibnamefont{Fayard}},
  \bibinfo{author}{\bibfnamefont{L.}~\bibnamefont{Henriet}},
  \bibinfo{author}{\bibfnamefont{A.}~\bibnamefont{Asenjo-Garcia}},
  \bibnamefont{and} \bibinfo{author}{\bibfnamefont{D.~E.} \bibnamefont{Chang}},
  \bibinfo{journal}{Phys. Rev. Research} \textbf{\bibinfo{volume}{3}},
  \bibinfo{pages}{033233} (\bibinfo{year}{2021}).

\bibitem[{\citenamefont{Ruostekoski and Javanainen}(1997)}]{JavanienLLBec}
\bibinfo{author}{\bibfnamefont{J.}~\bibnamefont{Ruostekoski}} \bibnamefont{and}
  \bibinfo{author}{\bibfnamefont{J.}~\bibnamefont{Javanainen}},
  \bibinfo{journal}{Phys. Rev. A} \textbf{\bibinfo{volume}{56}},
  \bibinfo{pages}{2056} (\bibinfo{year}{1997}).

\bibitem[{\citenamefont{Manzoni et~al.}(2017)\citenamefont{Manzoni, Chang, and
  Douglas}}]{Chang_OD}
\bibinfo{author}{\bibfnamefont{M.~T.} \bibnamefont{Manzoni}},
  \bibinfo{author}{\bibfnamefont{D.~E.} \bibnamefont{Chang}}, \bibnamefont{and}
  \bibinfo{author}{\bibfnamefont{J.~S.} \bibnamefont{Douglas}},
  \bibinfo{journal}{Nature Communications} \textbf{\bibinfo{volume}{8}},
  \bibinfo{pages}{1743} (\bibinfo{year}{2017}).

\bibitem[{\citenamefont{Ferioli et~al.}(2021)\citenamefont{Ferioli,
  Glicenstein, Henriet, Ferrier-Barbut, and Browaeys}}]{Antoine_memory}
\bibinfo{author}{\bibfnamefont{G.}~\bibnamefont{Ferioli}},
  \bibinfo{author}{\bibfnamefont{A.}~\bibnamefont{Glicenstein}},
  \bibinfo{author}{\bibfnamefont{L.}~\bibnamefont{Henriet}},
  \bibinfo{author}{\bibfnamefont{I.}~\bibnamefont{Ferrier-Barbut}},
  \bibnamefont{and} \bibinfo{author}{\bibfnamefont{A.}~\bibnamefont{Browaeys}},
  \bibinfo{journal}{Phys. Rev. X} \textbf{\bibinfo{volume}{11}},
  \bibinfo{pages}{021031} (\bibinfo{year}{2021}).

\bibitem[{\citenamefont{Kumlin et~al.}(2020)}]{Buchler_Topo}
\bibinfo{author}{\bibfnamefont{J.}~\bibnamefont{Kumlin}} \bibnamefont{et~al.},
  \bibinfo{journal}{Phys. Rev. A} \textbf{\bibinfo{volume}{102}},
  \bibinfo{pages}{063703} (\bibinfo{year}{2020}).

\bibitem[{\citenamefont{Perczel et~al.}(2017)}]{Topo_Lukin}
\bibinfo{author}{\bibfnamefont{J.}~\bibnamefont{Perczel}} \bibnamefont{et~al.},
  \bibinfo{journal}{Phys. Rev. Lett.} \textbf{\bibinfo{volume}{119}},
  \bibinfo{pages}{023603} (\bibinfo{year}{2017}).

\bibitem[{\citenamefont{Williamson et~al.}(2020)\citenamefont{Williamson,
  Borgh, and Ruostekoski}}]{WG_Blockade}
\bibinfo{author}{\bibfnamefont{L.~A.} \bibnamefont{Williamson}},
  \bibinfo{author}{\bibfnamefont{M.~O.} \bibnamefont{Borgh}}, \bibnamefont{and}
  \bibinfo{author}{\bibfnamefont{J.}~\bibnamefont{Ruostekoski}},
  \bibinfo{journal}{Phys. Rev. Lett.} \textbf{\bibinfo{volume}{125}},
  \bibinfo{pages}{073602} (\bibinfo{year}{2020}).

\bibitem[{\citenamefont{Förster}(1948)}]{Foester}
\bibinfo{author}{\bibfnamefont{T.}~\bibnamefont{Förster}},
  \bibinfo{journal}{Annalen der Physik} \textbf{\bibinfo{volume}{437}},
  \bibinfo{pages}{55} (\bibinfo{year}{1948}).

\bibitem[{\citenamefont{Zaghloul and Ali}(2012)}]{faddeeva1}
\bibinfo{author}{\bibfnamefont{M.~R.} \bibnamefont{Zaghloul}} \bibnamefont{and}
  \bibinfo{author}{\bibfnamefont{A.~N.} \bibnamefont{Ali}},
  \bibinfo{journal}{ACM Trans. Math. Softw.} \textbf{\bibinfo{volume}{38}}
  (\bibinfo{year}{2012}).

\bibitem[{\citenamefont{Asenjo-Garcia
  et~al.}(2019{\natexlab{b}})\citenamefont{Asenjo-Garcia, Kimble, and
  Chang}}]{Asenjo-Garcia2019}
\bibinfo{author}{\bibfnamefont{A.}~\bibnamefont{Asenjo-Garcia}},
  \bibinfo{author}{\bibfnamefont{H.~J.} \bibnamefont{Kimble}},
  \bibnamefont{and} \bibinfo{author}{\bibfnamefont{D.~E.} \bibnamefont{Chang}},
  \bibinfo{journal}{Proceedings of the National Academy of Sciences}
  \textbf{\bibinfo{volume}{116}}, \bibinfo{pages}{25503}
  (\bibinfo{year}{2019}{\natexlab{b}}).

\bibitem[{\citenamefont{Shi et~al.}(2015)\citenamefont{Shi, Chang, and
  Cirac}}]{Tao2015}
\bibinfo{author}{\bibfnamefont{T.}~\bibnamefont{Shi}},
  \bibinfo{author}{\bibfnamefont{D.~E.} \bibnamefont{Chang}}, \bibnamefont{and}
  \bibinfo{author}{\bibfnamefont{J.~I.} \bibnamefont{Cirac}},
  \bibinfo{journal}{Phys. Rev. A} \textbf{\bibinfo{volume}{92}},
  \bibinfo{pages}{053834} (\bibinfo{year}{2015}).

\end{thebibliography}

\section*{Methods} 
\noindent
\textbf{Data aquisition}\newline
For the detection of the transmitted light we use an InGaAs single-photon detector in combination with a time tagger.
Since a large fraction of the guided mode (76$\%$) that is not interacting with atoms creates a strong background signal, we employ a virtual lock-in amplification:  The 780 nm pump laser is amplitude modulated at 10 kHz using an acousto-optic modulator. That modulated signal is also sent to the time tagger and synchronized with the detection channel. By demodulating the transmitted probe we can reduce the noise level of the signal and suppress the strong background light level.
This technique allows us to evaluate background-free transmission signals at probe powers on the fW scale in the telecom wavelength regime.\newline

\noindent
\textbf{Density}\newline
The measured density is determined from the empirical equation for the thermal vapor pressure \cite{Steck}. We operate the pump laser well above saturation and thereby assume that 50$\%$ of the ground state population is excited to the intermediate 5P$_{3/2}$ state. 
\newline

\begin{figure}[h]
    \centering
    \includegraphics[width = \columnwidth]{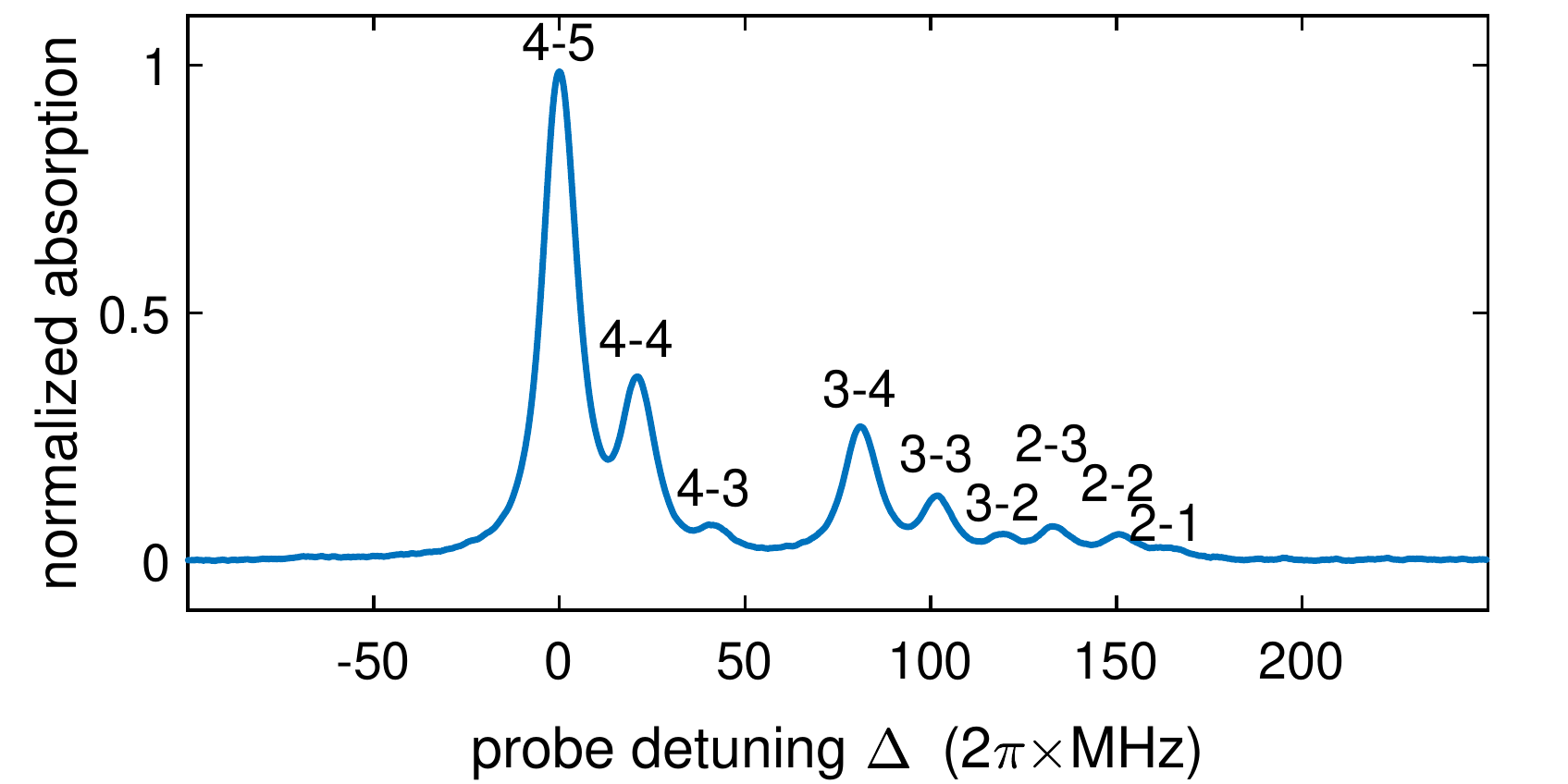}
    \caption{\textbf{Hyperfine reference spectrum of the 5S$_{1/2}$$\rightarrow$5P$_{3/2}$$\rightarrow$4D$_{5/2}$ two-photon transitions}.
    The numbers above each peak correspond to the hyperfine levels of the 5P$_{3/2}$ and 4D$_{5/2}$ state, respectively. The position of the cyclic transition (4-5) matches with the dashed line in Fig.~\ref{fig:fig2}e-f). Due to the inverted level structure of the 4D$_{5/2}$ state, smaller hyperfine levels are at higher energies.}
    \label{fig:supp1}
\end{figure}

\noindent
\textbf{Shift measurement}\newline
Together with the nanodevice cell, we use a 10 cm-long Rb reference cell at a fixed temperature of 40°C. Counterpropagating pump and probe lasers are sent through this vapor cell and by scanning the probe laser we obtain a reference signal that consists of 9 hyperfine transitions from the 5P$_{3/2}$ F = 2,3,4 states to the dipole allowed 4D$_{5/2}$ states with an average linewidth of $\Gamma_0/(2\pi) \approx 2$ MHz as depicted in Fig.~\ref{fig:supp1}. Despite locking the pump laser to the 5S$_{1/2}$ F = 3 to 5P$_{3/2}$ F = 4 transition we also excite the 5P$_{3/2}$ F = 2 and F = 3 states due to the Doppler detuning of thermal atoms.
This reference cell signal is used to measure lineshifts of the waveguide transmission signal. Since we resolve all dipole-allowed 4D$_{5/2}$ hyperfine states we can extract the type (attractive or repulsive) of dipolar interaction in the slot waveguide by comparing the measured shift to the individual hyperfine transitions.
\newline

\noindent
\textbf{Data evaluation}\newline
There are two parameters that are not captured by the simulation: The coupling efficiency to the grating couplers and an overall negative offset caused by the single-body Casimir-Polder effect. Latter, shifts all measured data points in Fig.~\ref{fig:fig2} by $\approx -70\Gamma_0$. We can measure the absolute transmitted power on a fW scale but due to uncertainties in the coupling efficiency the actual power within the slot is not exactly known. This causes an overall shift on the logarithmic intensity axis. We account for both of these uncertainties by fitting the simulated traces to the measured data with the single-body redshift and a global intensity factor as the only free fit parameters. The data in Fig.~\ref{fig:fig2} and Fig.~\ref{fig:fig3} is corrected for those two global offsets.
\newline

\begin{figure}[h]
    \centering
   \includegraphics[width = \columnwidth]{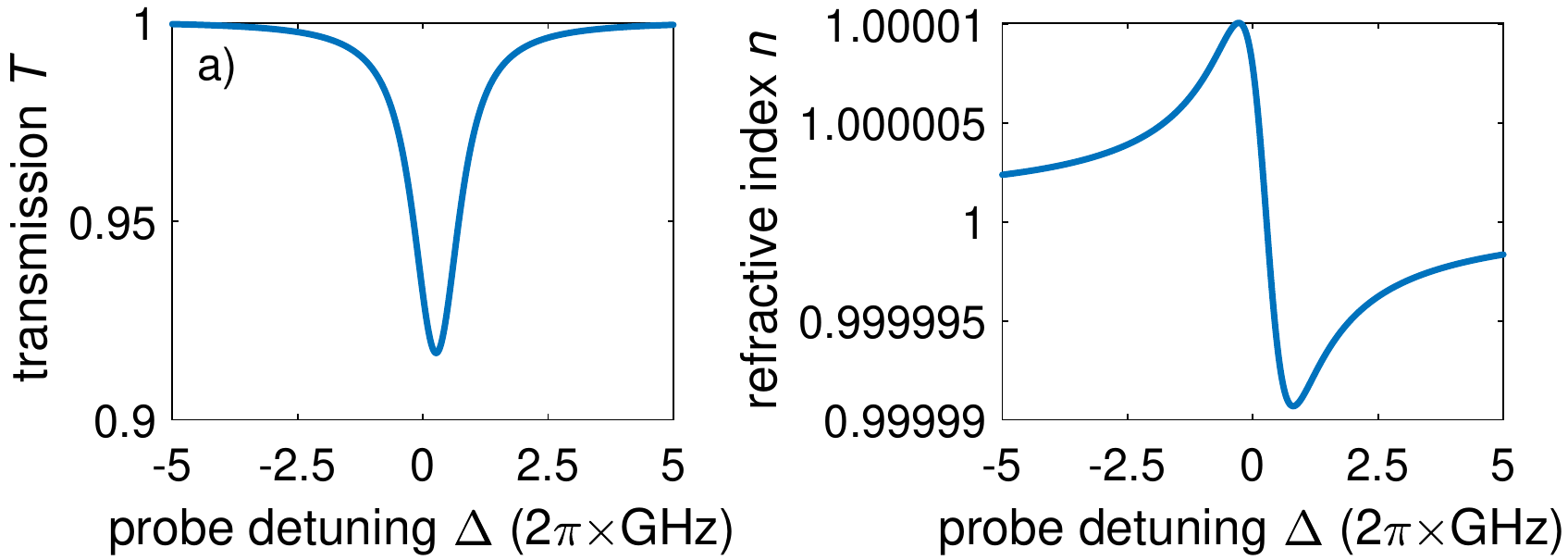}
    \caption{\textbf{Transmission data and the extracted refractive index spectrum}.
    a) Fitted transmission spectrum at the lowest measured intensity at $(kr)^{-3} = 6.7$. b) Refractive index extracted from the transmission spectrum in a).}
    \label{fig:supp4}
\end{figure}

\noindent
\textbf{Kerr nonlinearity}\newline
In order to estimate a value for the Kerr nonlinearity $n_2$ we first extract the refractive index from the transmission spectrum of the slot waveguide in the weak probe regime. We obtain the imaginary part of the susceptibility $\chi$ from the transmission $T = e^{-n_\text{eff}kL\text{Im}(\chi)}$ with the wavevector $k = 2\pi/\lambda$, $n_\text{eff}=2.53$, and the length of the interaction region $L = 200~\mu$m. The real part of the susceptibility can be calculated from the Kramers-Kroning relation. With this we obtain the refractive index from $n = \sqrt{1+\text{Re}(\chi)}$. The transmission spectrum and the extracted refractive index can be seen as a function of probe detuning in the extended data Fig.~\ref{fig:supp4}. We use the following expression~\cite{c17ms} to calculate the Kerr coefficient\\
\begin{equation*}
    n_2 = \frac{s}{I}\left|\frac{dn}{d\Delta}\right|,
\end{equation*}
where $s = 100$~MHz describes the lineshift of the spectrum, $I = 22I_\text{sat}$ the required intensity and $|dn/d\Delta| = 3.09\times10^{-8}$~MHz$^{-1}$ is the slope of the refractive index on resonance. We obtain $n_2 = 1.98\times10^{-7}$~m$^2$/W, which is about 2 orders of magnitude smaller than the value reported for electromagnetically induced transparency (EIT) in ultracold sodium~\cite{c17ms}. Note, that we used the probe intensity in the equation above since the interactions can be reduced to an effective two-level system addressed by the probe laser whereas the nonlinearity of the ultracold EIT system stems from the AC stark shift created by the pump. Now that we have the Kerr coefficient we can calculate the phase shift created by a single photon\\
\begin{equation}
    \phi = k_\text{eff}Ln_2I_\text{photon}.
\end{equation}
The intensity of a single photon can be estimated from\\
\begin{equation}
    I_\text{photon} = \frac{P_\text{photon}}{A_\text{eff}} = \frac{\hbar\omega}{TA_\text{eff}},
\end{equation}
with the transition frequency $\omega = 2\pi c/\lambda$, the inverse bandwidth of a photon that can resolve the lineshift $T = 1/s$ and the effective mode area $A_\text{eff}$. We obtain $A_\text{eff}$ from COMSOL simulations by integrating the Poynting vector $S$ over the cross section $dA$\\
\begin{equation}
    A_\text{eff} = \frac{1}{\text{max}(\textbf{S})}\int \textbf{S}d\textbf{A} = 7.68\times10^{-14}~\text{m}^2.
\end{equation}
For the calculation of $n_2I_\text{photon}(\text{Na})$ we used $n_2 = 1.80\times10^{-5}$~m$^2$/W~\cite{c17ms} and $I_\text{photon}$ was calculated with $T = 1/(1.30~\text{MHz})$ and $A_\text{eff} = 1.77\times10^{-10}$~m$^2$. As for the estimation of $n_2I_\text{photon}(\text{Rb})$ we used the reported value for the single photon phase shift $\phi\approx 2\times10^{-5}$~rad~\cite{EIT_RbXPS} in a 1.4~mm long sample probed at 780.24~nm.
\newline

\noindent
\textbf{Data fitting}\newline
As shown in previous studies~\cite{Artur}, the asymmetry caused by the Casimir-Polder potential is negligibly small for the broadened spectrum. Moreover, at the measured densities the lineshape is not significantly affected by the dipole-dipole interactions. In order to capture the lineshape of the signal across the whole intensity range consistently, we use a Voigt profile modified with a Fano lineshape as
\begin{equation}
    f(\Delta-s) = \frac{2\sqrt{\ln{2}}A}{(q^2-1)\omega_\text{D}\sqrt{\pi}}\left[ (q^2-1)\Re{(w)} -2q\Im{(w)}\right].
    \label{fit}
\end{equation}
Here, $w(\Delta-s)$ is given by the complex Faddeeva function ~\cite{faddeeva1}, $A$ is the amplitude, $s$ is the lineshift and $q$ is an asymmetry parameter obtained from the Fano lineshape. The asymmetry vanishes for $q\rightarrow \infty$. The Doppler broadening is given by the cell temperature. We fit the shift, the amplitude, the asymmetry parameter, and the Lorentzian contribution to the Voigt lineshape. At high probe intensities $I\gg I_\text{sat}$ we notice Fano-like interference of the guided photons carrying the atom-light information with strong background reflections. This behavior is captured by the asymmetry parameter $q$ without affecting the lineshift. Note that simulated traces at densities exceeding the experimental conditions show lineshapes that cannot be captured by this asymmetry parameter.\newline

\noindent
\textbf{Spin Model\label{sec:model}}\newline
In this section we detail the spin model that has been used to capture the single-body dynamics of the atom-laser interaction as well as the many-body effect arising due to the dipole-dipole interaction, mediated via both the free-space as well as the  waveguide mode. To include both the dispersive as well as the dissipative dynamics we use the Lindblad mater equation to describe the time evolution of the joint density matrix of the atomic ensemble as ($\hbar = 1$)
\begin{equation}~\label{eq:master equation}
    \dot{\hat{\rho}}(t) = -i[\hat{H}_\textrm{F} + \hat{H}_\textrm{int} ,  \hat{\rho}(t)] + \mathcal{D}(\hat{\rho}(t))\, ,
\end{equation}
where $\hat{H}_\textrm{F}$ and $\hat{H}_\textrm{int}$ describe the free-atom and the interacting atom ensemble unitary evaluations, respectively.

Let us consider an ensemble of $N$ identical atoms with frequency $\omega_a$ that are coherently driven with a laser at frequency $\omega_L$. For $\Omega_m$ being the Rabi frequency of the $m^\textrm{th}$-atom, the Hamiltonian of the free atoms reads as
\begin{equation}~\label{eq:free H}
    \hat{H}_\textrm{F} = \Delta_0 \sum_{m=1} ^ N  \hat{\sigma}_{ee}^m - \left(\Omega_m \hat{\sigma}_{eg}^m + \Omega_m^* \hat{\sigma}_{ge}^m \right)\, ,
\end{equation}
where $\Delta_0 = \omega_a - \omega_L$ is the atom-laser detuning, $\hat{\sigma}_{eg,ge}^m$ are the $m^\textrm{th}$-atom rising and lowering operators, respectively, and $\hat{\sigma}_{ee}^m$ is the corresponding excited state projection operator. 
Here, it has been assumed implicitly that the number of photons are large enough that the laser field can be treated classically hence, the Bloch equations give a proper description of the system dynamics. Note that this assumption is also consistent with the Born approximation we use later for the waveguide fields where we ignore any back actions from the atom radiations on the waveguide mode.

Aside from the free-atom part, there is a photon-mediated excitation exchange between the atoms which in the electric-dipole approximation is given as~\cite{Asenjo-Garcia2019}  
\begin{equation}~\label{eq:interaction H}
    \hat{H}_\textrm{int} = -3\pi \Gamma_0 \sum_{m,n=1}^N \Vec{d}_m^*\cdot\textrm{Real}(G^{mn})\cdot\Vec{d}_n ~\hat{\sigma}_{eg}^{(m)} ~\hat{\sigma}_{ge}^{(n)}
    \, ,
\end{equation}
where $\Gamma_0$ is the natural decay rate of atoms, $G^{mn}$ is the Green's tensor of $m^\textrm{th}$-atom at the location of $n^\textrm{th}$-one, and vice versa, and $\Vec{d}_m$ is the transition dipole moment of the $m^\textrm{th}$-atom.

In addition to the coherent interaction the dipolar interaction is accompanied with some incoherent dissipative behavior in the Markovian limit given by the following dissipator form
\begin{align}~\label{eq:Lindblad term}
    \mathcal{D}(\hat{\rho}) = 3\pi \Gamma_0 \sum_{m,n=1}^N \Vec{d}_m^*\cdot \textrm{Imag}(G^{mn}) \cdot     \Vec{d}_n \\ \left(2 \hat{\sigma}_{ge}^{(m)} \hat{\rho} ~ \hat{\sigma}_{eg}^{(n)} - \{\hat{\sigma}_{eg}^{(n)} \hat{\sigma}_{ge}^{(m)} , \hat{\rho}\} \right)\, .\nonumber
\end{align}
In a homogeneous isotropic medium, the dyadic Green's tensor of an electric-dipole reads as
\begin{equation}~\label{eq:GFS}
    G_\text{FS}^{m,m'}= \left(\bar{\bar{I}}+\frac{1}{k^2}\nabla\nabla\right)\frac{e^{ik|\textbf{a}_m-\textbf{a}_{m'}|}}{4\pi k|\textbf{a}_m-\textbf{a}_{m'}|},
\end{equation}
where $\textbf{a}_m$ is the $m^\textrm{th}$-atom location, $\bar{\bar{I}}$ is the unity tensor, and $k$ is the wavenumber.

As can be seen, $\textrm{Real}(G_\textrm{FS}^{m,m})$, i.e. Green's function at dipole location, is divergent leading to an effective energy re-normalization. However, its imaginary part, describing the self-action of the radiated field on itself, is the finite value of $1/(6\pi)$. From Eq.~\ref{eq:Lindblad term} it is apparent that one gets the decay rate of $\Gamma_0/2$ as expected for the spontaneous emission rate of a free atom in vacuum. 

Although the complete treatment of the interaction requires the full dyadic Green's tensor as can be inferred from Eq.~\ref{eq:interaction H} and Eq.~\ref{eq:Lindblad term}, for the numerical results presented in this work we only consider the diagonal terms, corresponding to aligned atoms along the laser electric field. Note that ignoring the off-diagonal entries is consistent with our previous assumption about the strong driving laser.\newline

\noindent
\textbf{Mean-Field Approximation\label{subsec:MF}}\newline
The exponential growth of the Hilbert space size as $2^N$ for an $N$-atom ensemble renders the exact solution of the Lindblad master equation of Eq.~\ref{eq:master equation} impossible. In this section, we first derive the Heisenberg-Langevin equations of motion (EoM) for the spins and later employ the moment factorization approximation to obtain the mean-field (MF) EoM.

For an operator $\hat{\mathcal{O}}(t)$, the master equation can be employed to find the EoM as 
\begin{equation}~\label{eq:Langevin equation}
    \frac{d}{dt}\hat{\mathcal{O}}(t) = i\left[\hat{H} , \hat{\mathcal{O}} \right] +  \mathcal{D}(\hat{\mathcal{O}})\, .
\end{equation}
From Eq.~\ref{eq:interaction H} and Eq.~\ref{eq:Lindblad term} we have
\begin{align}~\label{eq:spin exchange}
    \hat{H} & = \Delta_0 \sum_{m=1}^N \hat{\sigma}_{ee}^{(m)} - \left(\Omega_m \hat{\sigma}_{eg}^{(m)} + \Omega_m^* \hat{\sigma}_{ge}^{(m)}\right)  \\  
    &   + \sum_{n \ne m} \frac{J_{mn}}{2} \left(\hat{\sigma}_{eg}^{(m)} \hat{\sigma}_{ge}^{(n)}  + \hat{\sigma}_{eg}^{(n)} \hat{\sigma}_{ge}^{(m)}\right)\, ,\nonumber
\end{align}
and 
\begin{equation}~\label{eq:Lindblad operator}
    \mathcal{D}(\hat{\mathcal{O}}) = \sum_{m,n=1}^N \frac{\Gamma_{mn}}{2} \left(2\hat{\sigma}_{eg}^{(n)} \hat{\mathcal{O}} \hat{\sigma}_{ge}^{(m)} - \{\hat{\sigma}_{eg}^{(m)} \hat{\sigma}_{ge}^{(n)} , \hat{\mathcal{O}}\}\right) \, ,
\end{equation}
where we defined $J_{mn} = -3\pi \Gamma_0 ~ \textrm{Real}(G^{mn})$ and $\Gamma_{mn}/2 = 3\pi \Gamma_0 ~  \textrm{Imag}(G^{mn})$ to write the operators in the symmetric form and remove the dispersive self-effect in re-normalizing the energy.

Using these forms we can obtain the spin EoM as 

\begin{widetext}~\label{eq:spin operator EoM}
\begin{align*}
    %
 %
%
 \frac{d}{dt} \hat{\sigma}_{ge}^{(k)} &= -i\Delta_0 \hat{\sigma}_{ge}^{(k)} - i\Omega_k \left(\hat{\sigma}_{ee}^{(k)} - \hat{\sigma}_{gg}^{(k)}\right) 
 + i\sum_{n \ne k=1}^N J_{nk} \left( \hat{\sigma}_{ge}^{(n)} \hat{\sigma}_{ee}^{(k)} -  \hat{\sigma}_{ge}^{(n)} \hat{\sigma}_{gg}^{(k)} \right) 
-\frac{\Gamma_0}{2} \hat{\sigma}_{ge}^{(k)}
+ \sum_{n \ne k} \frac{\Gamma_{nk}}{2} \left( \hat{\sigma}_{ge}^{(n)} \hat{\sigma}_{ee}^{(k)} -  \hat{\sigma}_{ge}^{(n)} \hat{\sigma}_{gg}^{(k)} \right)\, ,
\\
 \frac{d}{dt} \hat{\sigma}_{ee}^{(k)} &= i\left(\Omega_k \hat{\sigma}_{eg}^{(k)} - \Omega_k^* \hat{\sigma}_{ge}^{(k)} \right) + i\sum_{n \ne k=1}^N J_{nk} \left(\sigma_{eg}^{(n)} \sigma_{ge}^{(k)} - \hat{\sigma}_{ge}^{(n)} \hat{\sigma}_{eg}^{(k)} \right) - \Gamma_0 \hat{\sigma}_{ee}^{(k)} - \sum_{n \ne k=1}^N \frac{\Gamma_{nk}}{2} \left(\hat{\sigma}_{ge}^{(n)} \hat{\sigma}_{eg}^{(k)} + \hat{\sigma}_{eg}^{(n)} \hat{\sigma}_{ge}^{(k)} \right)\, .
\end{align*}
\end{widetext}
In the above forms we separated the self-terms from the mutual ones to emphasize the dipole-dipole interactions. As can be inferred from these Langevin EoM such interactions lead to a hierarchy of moments, in general.

As the first approximation, one can ignore the quantum correlations and employ the factorization approximation as $\braket{\hat{\sigma}^{(m)} \hat{\sigma}^{(n)}} \approx \braket{\sigma^{(m)}} \braket{\sigma^n}$ to simplify the spin EoM to their MF form as
\begin{widetext}~\label{eq:MF MoM FS}
\begin{align*}
    %
%
 \frac{d}{dt} \braket{\hat{\sigma}_{ge}^{(k)}} &= -i\Delta_0 \braket{\hat{\sigma}_{ge}^{(k)}} - i\Omega_k \braket{\hat{\sigma}_{z}^{(k)}} 
 - i \braket{\hat{\sigma}_{z}^{(k)}} \sum_{n \ne k=1}^N \left(-J_{nk} + i \frac{\Gamma_{nk}}{2} \right) \braket{\hat{\sigma}_{ge}^{(n)}} - \frac{\Gamma_0}{2} \braket{\hat{\sigma}_{ge}^{(k)}} \, ,
\\
 \frac{d}{dt} \braket{\hat{\sigma}_{ee}^{(k)}} &= i \Omega_k \braket{\hat{\sigma}_{eg}^{(k)}}  + i \braket{\hat{\sigma}_{eg}^{(k)}} \sum_{n \ne k=1}^N \left(- J_{nk} + i\frac{\Gamma_{nk}}{2} \right) \braket{\hat{\sigma}_{ge}^{(n)}} + c.c. - \Gamma_0 \braket{\hat{\sigma}_{ee}^{(k)}} \, ,
\end{align*}
\end{widetext}
where $\hat{\sigma}^{(m)}_z = \hat{\sigma}^{(m)}_{ee} - \hat{\sigma}^{(m)}_{gg}$ is the $m^\textrm{th}$-atom inversion operator. 

Note that the above equations resemble a two-level atom Bloch equations if the $m^\textrm{th}$-atom Rabi frequency is replaced by an \emph{effective} one as

\begin{align}~\label{effective Rabi freq}
    \Omega_\text{eff}^m & = \Omega_0^m + \sum_{m \ne n = 1}^N \left(-J_{mn} + i\frac{\Gamma_{mn}}{2} \right) \braket{\hat{\sigma}_{ge}^{(n)}} \nonumber \\
    & = \Omega_0^m + 3\pi\sum_{m\neq n = 1} G^{m,n} \braket{\hat{\sigma}^{(n)}_{ge}} \, .
\end{align}
Recognizing that $\braket{\hat{\sigma}_{ge}^{(n)}}$ is related to the induced dipole of the $n^\textrm{th}$-atom, the last equation has a simple interpretation: within the mean-field treatment the effective field at the $m^\textrm{th}$-atom position is the summation of the incident laser field, i.e. $\Omega_0^m$, and the field radiated by all other atoms whose contributions are determined via $G^{mn}$, i.e. the electric field of $n^\textrm{th}$-atom at location of the $m^\textrm{th}$-one. Note that the complex effective Rabi frequency captures both the collective dispersive and dissipative effects. 

\begin{figure}[h]
    \centering
    \includegraphics[scale = 0.6]{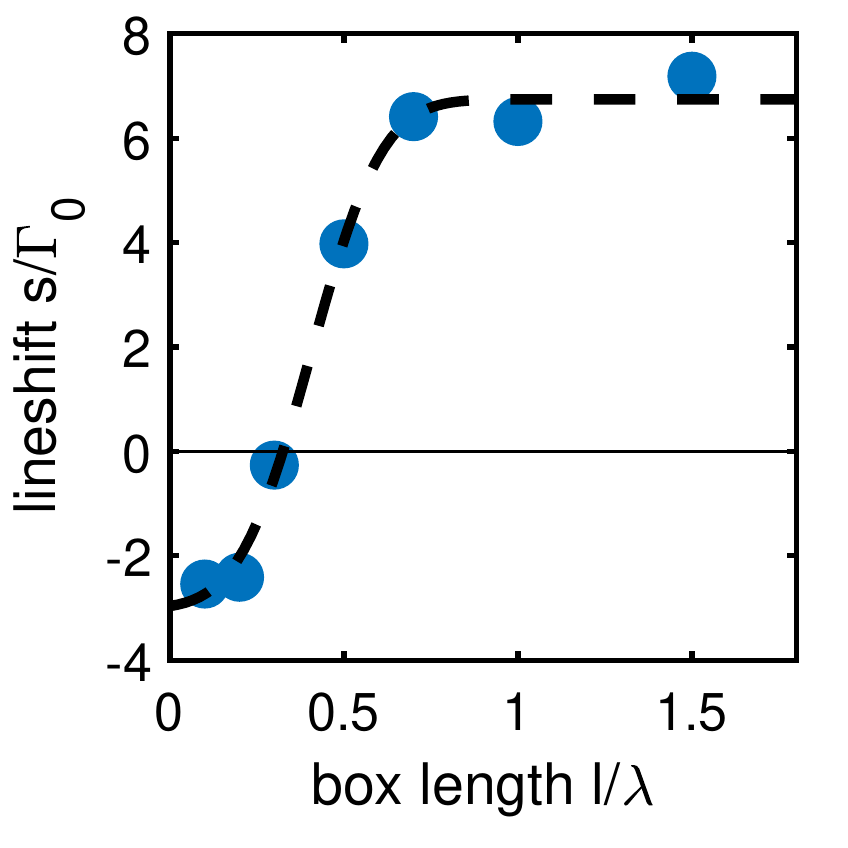}
    \caption{\textbf{Transition from 2D to pseudo 1D dipolar interactions}.
    Dipolar interaction induced linshift as a function of the cloud geometry, from a 2D to a pseudo 1D atomic cloud. Blue points show the simulated data as a function of the box length in a probe volume (box) of $(\Delta x, \Delta y, \Delta z) = (0.3\lambda, 0.3\lambda, l)$. The density is kept constant in all cases. The dashed line is as a guide to the eye.}
    \label{fig:supp2}
\end{figure}

Finally, the self-effect of the atom radiated field is simply described using the following dissipator
\begin{equation}~\label{eq:decoupled Lindblad}
    \mathcal{D}(\hat{\rho}) = \frac{\Gamma_0}{2}\sum_{m=1}^N \left(2\hat{\sigma}_{ge}^{(m)} \hat{\rho} ~ \hat{\sigma}_{eg}^{(m)} - \{\hat{\sigma}_{eg}^{(m)} \hat{\sigma}_{ge}^{(m)}, \hat{\rho} \} \right)\, .
\end{equation}
Fig.~\ref{fig:supp2} shows the simulated lineshift of an atomic ensemble in free space confined in a simulation box of $(\Delta x, \Delta y, \Delta z) = (0.3 \lambda, 0.3\lambda, l)$ as a function of the box length $l$, and driven with a uniform weak laser field along the $z$-direction. By changing the box length while keeping $\Delta x , \Delta y$ fixed, the cloud shows a transition from a 2D to a pseudo 1D case. In the 2D scenario, where $l \ll \lambda$ attractive interactions dominate and we obtain a redshift of the spectrum. As we increase the third dimension we cross the zero line when all dimensions are equal in size. Further increase in the simulation box length leads to a pseudo 1D case where $l>\lambda$. In this regime mainly repulsive dipolar interaction contribute to the lineshift of the spectrum. 

As shown in this sub-section the semi-classical MF approximation of an interacting atomic ensemble is reduced to the system Green's function determination. 
While for the atoms in a homogeneous medium one can use the analytic dyadic Green's tensor in Eq.~\ref{eq:GFS}, there only a handful of other problems that a closed-form Green's function is attainable. For the other problems, including the slot waveguide in our case, the exact determination of the Green's tensor relies on numerical calculations. Due to the strong confinement of the mode inside the slot-waveguide however, its Green's function can be approximated by some corrections to the free-space case, as will be discussed in the next sub-section. 
\newline

\noindent
\textbf{Waveguide-Mediated Interaction\label{subsec:WG}}\newline
In order to derive an effective Green's function for the slot wavgeuide, in this section we employ the input-output formalism to describe the dynamics of a 1D spin chain coupled to a waveguide~\cite{Tao2015}. Note that due to the tight confinement of the waveguide mode inside the deep sub-wavelength slot ($w \ll \lambda$) the 1D spin chain is a proper approximation of our system.

Consider a 1D chain of $N$ atoms at $z_{i=1,\cdots, N}$ coupled to a waveguide along the $z$-direction. For each propagation constant of $\beta$ there are two modes, a forward propagating mode ($\hat{a}_R(\beta)$) varying as $e^{i \beta z}$ and a backward-propagating one ($\hat{a}_L(\beta)$) as $e^{-i\beta z}$. Within the validity range of RWA, i.e. a separate energy scale between the detuning and the resonance frequency of the atoms ($\Delta_0 \ll \omega_a$), the unitary dynamics of the atom-mode coupling is given as follows
\begin{widetext}~\label{eq:chain waveguide Hamiltonian}
\begin{align}
\hat{H} = \int_0^\infty d\beta ~ \omega_{ph}(\beta) \left(\hat{a}_R^\dagger(\beta) \hat{a}_R(\beta) + \hat{a}_L^\dagger(\beta) \hat{a}_L(\beta) \right) + \omega_a \sum_{n=1}^N \hat{\sigma}_{ee}^{(n)}  - \sum_{n=1}^N g_0^{(n)} \int_0^\infty d\beta~ \hat{\sigma}_{eg}^{(n)} \left(\hat{a}_R(\beta) e^{i \beta z_n}+ \hat{a}_L(\beta) e^{-i \beta z_n} \right) + \textrm{H.c.},
\end{align}
\end{widetext}
where in the above equation $\omega_{ph}(\beta)$ is the waveguide dispersion and $g_0^{(n)}$ is the vacuum Rabi frequency of the $n^\textrm{th}$-atom-mode coupling. Due to the translational symmetry of the modes along the $z$-direction, $g_0$ only depends on the transverse coordinates of the atom. Furthermore, as we separated the forward and backward propagating modes explicitly, $\beta \ge 0$.

If a Markovian process describes the coupling of the atoms to non-guided modes with a rate of $\Gamma_1$, then we can describe this process with the following dissipator 
\begin{equation}~\label{eq:chain dissipator}
    \mathcal{D}(\hat{\rho}) = \frac{\Gamma_1}{2} \sum_{n=1}^N \left(2\hat{\sigma}_{ge}^{(n)} \hat{\rho} ~ \hat{\sigma}_{eg}^{(n)} - \{\hat{\sigma}_{ee}^{(n)} , \hat{\rho} \}\right) \, .
\end{equation}
Following a similar procedure as in the previous section, we can determine Langevin EoM for the waveguide modes and spins as follows
\begin{widetext}~\label{eq:WG modes solutions in the chain}
\begin{align}
    \dot{\hat{a}}_R(\beta) &= -i\omega_{ph}(\beta) \hat{a}_R(\beta) + i \sum_{n=1}^N g_0^{(n)} \hat{\sigma}_{ge}^{(n)} e^{-i \beta z_n} \, ,\\ \nonumber
    \dot{\hat{a}}_L(\beta) &= -i\omega_{ph}(\beta) \hat{a}_L(\beta) + i \sum_{n=1}^N g_0^{(n)} \hat{\sigma}_{ge}^{(n)} e^{+i \beta z_n} \, ,\\ \nonumber
    \dot{\hat{\sigma}}_{ge}^{(n)} &= -i\omega_a \hat{\sigma}_{ge}^{(n)} - i g_0^{(n)} \left(\hat{\sigma}_{ee}^{(n)} - \hat{\sigma}_{gg}^{(n)} \right) \int_0^\infty d\beta~ \left(\hat{a}_R(\beta) e^{i \beta z_n} + \hat{a}_L(\beta)e^{-i \beta z_n} \right) - \frac{\Gamma_1}{2} \hat{\sigma}_{ge}^{(n)} \, .
\end{align}
\end{widetext}
As can be seen, in the absence of atoms the first two equations for the waveguide modes decouple and reduce to the time harmonic evolution of the right and left propagating waves. The presence of the atoms however, couples these two modes as captured by the last equation. 

To find the equations of motion for the spin degrees of freedom only, hence the effective spin model, we can integrate the waveguide mode operators in terms of the spin operators. To arrive at a \emph{closed-form} for the spin dynamics, we need some assumptions for the waveguide dispersion relation, i.e. $\omega_{ph}(\beta)$. As there is not any dispersion engineering for the slot waveguide we use the linear dispersion approximation with a fixed phase velocity as $\omega_{ph}(\beta) = v_p \beta$. 

Using this approximation it is straightforward to show that the spins' dynamics satisfy the following equation
\begin{widetext}~\label{eq:spin solution2}
\begin{align}
    \dot{\hat{\sigma}}_{ge}^{(n)} &= -i\left(\omega_a -i \frac{\Gamma_1}{2} \right) \hat{\sigma}_{ge}^{(n)} - i g_0^{(n)} \sqrt{2\pi} \hat{\sigma}_z^{(n)} \times  \left[\left(\hat{a}_R^{(in)}(z_n - v_p t) + \hat{a}_L^{(in)} (z_n + v_p t )\right) + i \frac{\sqrt{2\pi}}{v_p} \sum_{m=1}^N g_0^{(m)} \hat{\sigma}_{ge}^{(m)} \left(t - \frac{|z_n - z_m|}{v_p} \right) \right]\, ,
\end{align}
\end{widetext}
where $\hat{a}_{R,L}^{(in)}$ are the input fields for the forward and backward propagating modes, respectively. 

Assume that at $t = 0$ we excite the $\beta_0$ waveguide mode only from the right, and with strong enough amplitude $E_0$ such that the field operator can be approximated with its MF as
\begin{equation}~\label{eq:input CW mode}
    \hat{a}_R^{(in)}(x,0) \approx E_0 e^{i \beta_0 z}\, .
\end{equation}
Moreover, due to the small waveguide size hence the negligible retardation effect, we can use the Markov approximation to simplify the delayed spin dynamics to an instantaneous one as
\begin{align}~\label{eq:spin CW approximation}
    \hat{\sigma}_{ge}^{(n)}\left(t - \frac{|z - z'|}{v_p} \right) &\approx 
    \hat{\sigma}_{ge}^{(n)}(t) + \dot{\hat{\sigma}}_{ge}^{(n)}(t)\frac{|z - z'|}{v_p} \nonumber\\
    & \approx \hat{\sigma}_{ge}^{(n)}(t) \left(1 + \frac{\omega_0 |z - z'|}{v_p} \right) \nonumber\\
    & \approx \hat{\sigma}_{ge}^{(n)}(t) e^{i\beta_0 |z - z'|} \, .
\end{align}

Finally, by re-writing the spin dynamics in the rotated frame of the laser, i.e. $\omega_0 = v_p \beta_0$, as $\hat{\sigma}_{ge}^{(n)}(t) = \hat{\Tilde{\sigma}}_{ge}^{(n)}(t) e^{-i\omega_0 t}$, we get the following EoM for the slowly varying envelopes of the spins as
%

\begin{align}~\label{eq:atom-WG CW}
    \dot{\hat{\tilde{\sigma}}}_{ge}^{(n)} &= 
    -i\left(\delta_{a0} -i \frac{\Gamma_1}{2} \right) \hat{\Tilde{\sigma}}_{ge}^{(n)} - i g_0^{(n)} \sqrt{2\pi} \hat{\Tilde{\sigma}}_z^{(n)} \times \\
& \left[ E_0 e^{i \beta_0 z_n}  + i\frac{\sqrt{2\pi}}{v_p} \sum_{m=1}^N g_0^{(m)} \hat{\Tilde{\sigma}}_{ge}^{(m)} e^{i\beta_0 |z_n - z_m|} \right],
\end{align}

where in the above equation $\delta_{a0} = \omega_a - \omega_0$ is the atom-laser detuning.

As can be seen, this equation resembles a lot to the EoM of polarized atoms in a homogeneous medium where the atom-light interaction is treated classically via Maxwell's equations. In particular, we can recognize the effect of other atoms appearing in the summation, as an effective Lorentz field correction modifying the incident laser field, i.e. $E_0 e^{i \beta_0 z_n}$. 

To develop a better understanding of this final form and obtain the corresponding spin model similar to Eq.~\ref{eq:interaction H}, we separate the atom-atom interaction to determine the interaction Hamiltonian by inspection
\begin{widetext}
\begin{align}~\label{eq:atom-WG-CW2}
    \dot{\hat{\tilde{\sigma}}}_{ge}^{(n)} &= 
    -i\left(\delta_{a0} -i \frac{\Gamma_1}{2} \right) \hat{\Tilde{\sigma}}_{ge}^{(n)} - i \hat{\Tilde{\sigma}}_{z}^{(n)} \times \left(g_0^{(n)} \sqrt{2\pi} E_0 e^{i \beta_0 z_n}  + i\frac{2\pi g_0^{(n)^2}}{v_p} ~ \hat{\Tilde{\sigma}}_{ge}^{(n)} \right) 
    - i \hat{\Tilde{\sigma}}_{z}^{(n)} \sum_{m\ne n = 1}^N \frac{i 2 \pi g_0^{(n)} g_0^{(m)}}{v_p} e^{i\beta_0|z_n - z_m|} \hat{\Tilde{\sigma}}_{ge}^{(m)} \, .
    \end{align}
\end{widetext}
When compared with the interacting ensemble dynamics in Eq.~\ref{effective Rabi freq} we can deduce the effective 1D Green's function as
\begin{align}
    G_\textrm{WG}^{m,m'} & = i \frac{\Gamma_\textrm{1D}^{m,m'}}{\Gamma_0} e^{i\beta_0|z_m - z'_m|}\, ,\label{eq:1D WG GF}
\end{align}
where the waveguide coupling rate $\Gamma_\textrm{1D}^{m,m'}$ is defined as
\begin{equation}
    \Gamma_\textrm{1D}^{m,m'} = \frac{2 g_0^{(m)} g_0^{(m')}}{3 v_p}\, .
\end{equation}
The physical meaning of the above equation is rather simple: if $g_0$ is the radiation of one dipole coupled to the waveguide then the generated guided mode interacts with another atom with the same rate hence the effective interaction rate between the atoms should be proportional to $g_0^{(m)} g_0^{(m')}$. Moreover, this is related to the avaialble density of states related to the group velocity as $1/v_p$, and finally the phase delcay between the atoms set by $e^{i\beta_0 |\Delta z|}$. 

As mentioned before the atom radiation may couple to the non-guided modes with the rate of $\Gamma_1$, a parameter whose exact value requires an exact calculation of the Green's function. Here, we approximate the total Green's function of the slot waveguide perturbatively as
\begin{equation}~\label{eq:total-GF-FS+WG}
    G_\textrm{total}^{m,m'} \approx G_\textrm{FS}^{m,m'} + G_\textrm{WG}^{m,m'}\, .
\end{equation}
This form allows us to capture both the dominant near-field effect due to the free-space Green's function as well as infinite-range waveguide-mediated interaction via $G_\textrm{WG}^{m,m'}$. With this approximation we can employ the MF EoM of the previous section, where the Green's function is now replaced with $G_\textrm{total}^{m,m'}$. \newline

\noindent
\textbf{Clarifying Case: Single Atom Coupled to the Waveguide}\newline
To better illustrate the the waveguide modification of the atomic radiation it is helpful to consider the single-atom case where the effective Rabi frequency from Eq.~\ref{eq:atom-WG-CW2} is
\begin{equation}~\label{eq:WG effective Rabi}
    \Omega_\textrm{eff} = g_0 \sqrt{2\pi} E_0 + i \frac{2\pi g_0^2}{v_p} \braket{\hat{\tilde{\sigma}}_{ge}} \, .
\end{equation}
This modified rate is the coherent sum of the incident driving Rabi frequency (the first term) and the scattered field of the atom coupled back to the waveguide (the second term). A comparison with the waveguide-coupled results sheds the light on the physical implication of the scattered term in $\Omega_\textrm{eff}$ in Eq.~\ref{eq:WG effective Rabi} as $i 3\pi \Gamma_\textrm{1D} \braket{\hat{\tilde{\sigma}}_{ge}}$. In general, this is a complex quantity that modifies the Rabi flopping rate as well as the coherence decay rate. In the perturbative limit of the non-depleted atom however, i.e. $\braket{\hat{\sigma}_z} \approx -1$, the spin dynamics reads as
\begin{equation}
    \dot{\hat{\tilde{\sigma}}}_{ge} \approx 
     -i \delta_{a0} \hat{\Tilde{\sigma}}_{ge} - \left(\frac{\Gamma_1}{2} + 3\pi \Gamma_\textrm{1D}\right) \hat{\Tilde{\sigma}}_{ge} 
    +i g_0 \sqrt{2\pi} E_0 \, .
\end{equation}
When combined with the approximated Green's function of Eq.~\ref{eq:total-GF-FS+WG}, this leads to the total decay rate of
\begin{equation}~\label{eq:waveguide-mediated-decay}
    \Gamma_D = \Gamma_0 + \frac{4\pi g_0^2}{v_p} \, , 
\end{equation}
where $\Gamma_0$ is the free space decay rate.

On the other hand, in this non-depleted regime one can employ Fermi's golden rule to estimate the waveguide effect in modifying the decay rate as follows
\begin{equation}~\label{eq:Fermi's golden rule}
    \Gamma_\textrm{WG} = 2\pi |h_{eg}|^2 \mathcal{G}(\omega_a)\, ,
\end{equation}
where $|h_{eg}|$ is the coupling rate, i.e. $g_0$ in this case, and $\mathcal{G}(\omega_a)$ is the local density of the optical states (LDOS) at the atom's position and the transition frequency of $\omega_a$. 

Since the waveguide LDOS is inversely proportional to the phase velocity $v_p$ we get
\begin{equation}~\label{eq:waveguide decay}
    \Gamma_\textrm{WG} = 2\times 2\pi \frac{g_0^2}{v_p} = 4\pi \frac{g_0^2}{v_p} \, ,
\end{equation}
where the pre-factor 2 is due to the degeneracy of the left/right propagating modes. That means the total decay rate of the atom is $\Gamma_0 + \Gamma_\textrm{WG}$ which is identical to $\Gamma_D$ in Eq.~\ref{eq:waveguide-mediated-decay} obtained from the effective Bloch equations. Finally, one can determine the Purcell factor (PF) due to the waveguide coupling as
\begin{equation}
    \textrm{PF} = \frac{\Gamma_D}{\Gamma_0} = 1 + 6\pi \frac{\Gamma_\textrm{1D}}{\Gamma_0} \, .
\end{equation}

\begin{figure}[h]
    \centering
   \includegraphics[width = \columnwidth]{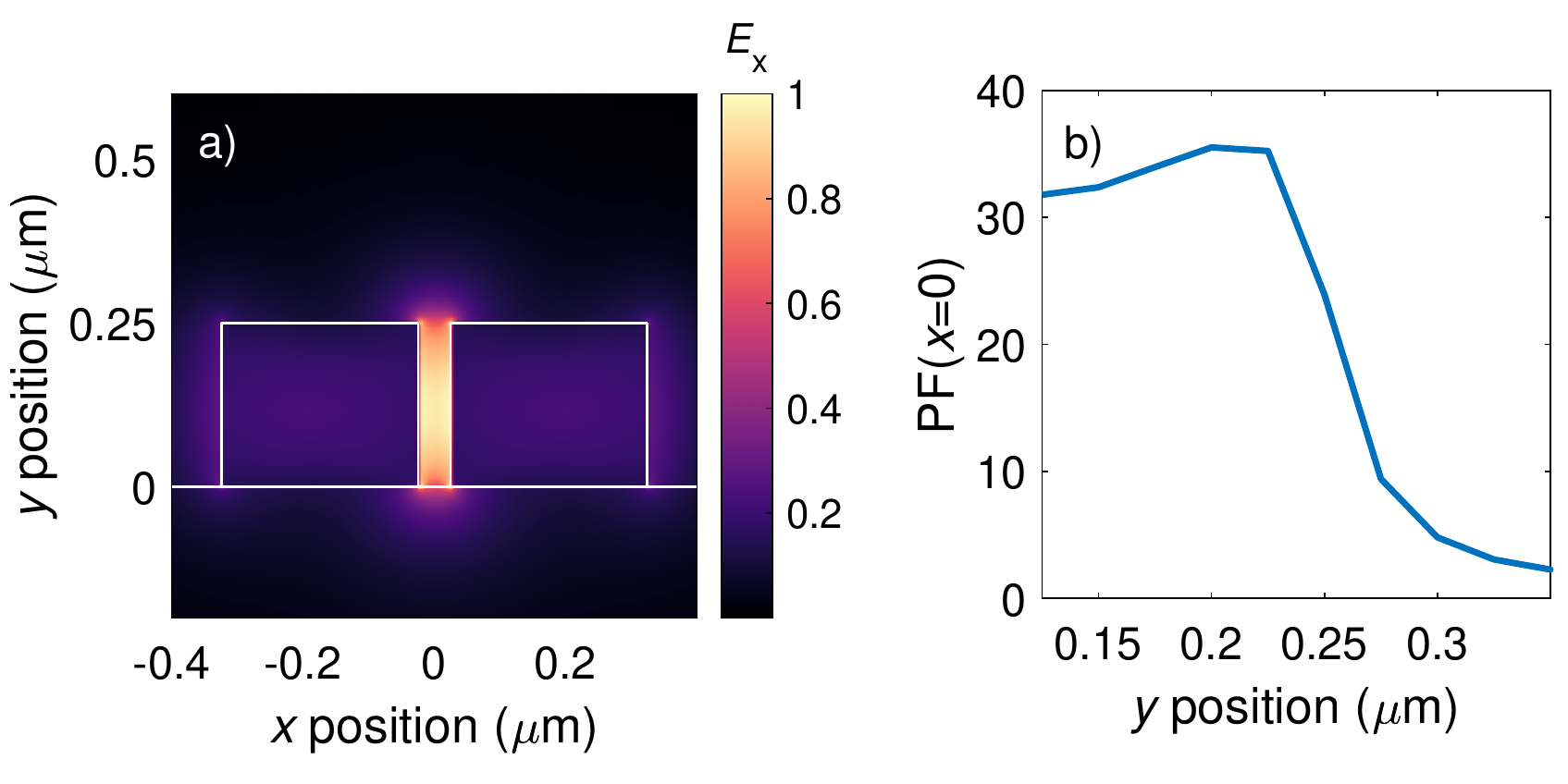}
    \caption{\textbf{Mode profile and Purcell factor}.
    a) Normalized electric field profile of the slot waveguide mode. Due to the sub-wavelength geometrical features the electric field  is dominantly $x$-polarized such that the components along $y$- and $z$-directions are almost negligible. White lines mark the silhouette of the waveguides. b) Purcell factor PF at the center of the waveguide and as a function of the vertical position $y$.}
    \label{fig:supp3}
\end{figure}

Fig.~\ref{fig:supp3} shows the cross section of the electric field profile of the slot waveguide and its corresponding Purcell factor calculated at the center of the slot where the electric field hence $g_0$ in the maximum. Right at the center a Purcell factor as high as $\approx 35$ can be achieved that decays almost exponentially away from the waveguide following the mode profile. 

Furthermore, by solving Eq.~\ref{eq:atom-WG-CW2} in the weak-probe limit the scattering parameters, i.e. the reflected $\braket{\hat{a}_L(z)}$ and the transmitted $\braket{\hat{a}_R(z)}$ modes can be determined as follows
\begin{widetext}
\begin{align}
    \braket{\hat{\Tilde{\sigma}}_{ge}}_\textrm{ss} & = \frac{ig_0 \sqrt{2\pi} E_0 }{i\delta_{a0} + \textrm{PF} \frac{ \Gamma_0}{2}}  \,, \\
    \braket{\hat{\tilde{a}}_L(z)}_\textrm{ss}  & = i\frac{g_0 \sqrt{2\pi}}{v_p} \braket{\hat{\Tilde{\sigma}}_{ge}}_\textrm{ss} e^{- i\beta_0 z} = - \frac{\Gamma_\textrm{WG} }{i2\delta_{a0} + \textrm{PF} \times \Gamma_0 } E_0 e^{-i\beta_0 z}\, , \\
    \braket{\hat{\tilde{a}}_R(z)}_\textrm{ss}  & = E_0 e^{i\beta_0 z} + i\frac{g_0 \sqrt{2\pi}}{v_p} \braket{\hat{\tilde{\sigma}}_{ge}}_\textrm{ss} e^{i\beta_0 z } = E_0 e^{i\beta_0 z} \left(1 - \frac{\Gamma_\textrm{WG}}{i2\delta_{a0} + \textrm{PF} \times \Gamma_0}\right)\, .
\end{align}
\end{widetext}
Notice the general Lorentzian lineshape of these quantities with a broadening of $\textrm{PF} \times \Gamma_0$ and centered at $\delta_{a0} = 0$, corresponding to a resonance excitation. As expected for a fixed coupling rates, the resonance is where the maximum reflection and the minimum transmission is expected. As this is the perturbative limit solution, there is not any power broadening hence the saturation, due to the large $E_0$. We would like to emphasize that, this is only the weak-probe limit and in our numerical simulations we used the complete form including the population inversion dynamics and saturation for a uniform treatment of the waveguide effect within the whole power range.\\
\noindent



\hfill
\break 

\end{document}